\documentclass[pre,onecolumn,longbibliography]{revtex4-1}
\usepackage{graphicx}

\begin{document}

\title{Laminar flow of two miscible fluids in a simple  network}
\date{\today}
\author{Casey M. Karst}
\author{Brian D. Storey}
\email[Address correspondence to: ]{brian.storey@olin.edu}
\author{John B. Geddes}
\address{Franklin W. Olin College of Engineering, Needham MA 02492}

\begin{abstract}

When a fluid comprised of multiple phases or constituents flows
 through a network,  non-linear phenomena such as multiple stable equilibrium states and
 spontaneous oscillations can occur. Such behavior has been observed
or predicted in a number of  networks including
  the flow of blood through the microcirculation, the flow of picoliter
droplets through microfluidic devices, the flow of magma through lava tubes, and two-phase flow
in refrigeration systems.
While the existence of non-linear phenomena in a network with many inter-connections containing
fluids with complex rheology may seem unsurprising, this paper demonstrates that
 even simple networks containing Newtonian fluids in laminar flow can demonstrate multiple equilibria.

The paper describes  a  theoretical and experimental investigation of
the laminar flow of two miscible Newtonian fluids of different density and  viscosity through a simple network.
The fluids stratify due to gravity and remain as nearly distinct phases with some mixing  occurring only by diffusion. This fluid system has the advantage that it is easily controlled and modeled, yet contains the key ingredients for network non-linearities. Experiments and  3D simulations are first used to explore how phases distribute at a single T-junction.
 Once the phase separation at a single junction is known, a  network model is developed which predicts multiple equilibria in the simplest of networks. The existence of multiple stable equilibria is confirmed experimentally and a criterion for   existence  is developed. The network results are generic and could be applied to or found in different physical systems.

\end{abstract}

\maketitle

\section{Introduction}
When a fluid comprised of multiple phases or constituents flows through a connected fluidic network,
it has been observed in different
 applications that the phase distribution may exhibit unsteady or non-unique flow for fixed inlet conditions.
 Such heterogeneous distribution  of phase within
network flows has been studied  at a variety of scales.
At the micro-scale, the flow of droplets or bubbles through microfluidic networks  can
 demonstrate bistabilty and spontaneous oscillations
 \cite{Jousse2006,Schindler2008,Fuerstman2007,Prakash2007, Joanicot2005}.
 These network non-linearities  have been exploited by researchers who have demonstrated microfluidic memory, logic, and control devices  \cite{Fuerstman2007,Prakash2007, Joanicot2005}.
 On the macro-scale,  models of magma flow with either temperature-dependent viscosity \cite{Helfrich1995}  or volatile-dependent viscosity \cite{Wylie1999} have shown the existence of multiple solutions on the pressure-flow curve which can lead to spontaneous oscillations.

Another network that can exhibit complex behavior is micro-vascular blood flow. Nobel prize winner August Krogh noted the heterogeneity of blood flow in the webbed feet of frogs in the early 1920's~\cite{Krogh:1921aa}. In the {\it Anatomy and Physiology of Capillaries} he wrote~\cite{Krogh:1922aa}
\begin{quote}
In single capillaries the flow may become retarded or accelerated from no visible cause; in capillary anastomoses the direction of flow may change from time to time.
\end{quote}
Numerous researchers have confirmed these observations over the years.
De Backer {\it et al.} reported  the first  visualization of micro-vascular  flow in critically ill patients and observed
alterations in red blood cell distribution in patients with sepsis ~\cite{debacker:2002}. Their  videos (in their supplementary materials) capture the heterogeneity of blood flow that Krogh referred to; spontaneous changes in flow speed and spontaneous change in the direction of flow in loops.

The distribution of red blood cells in micro-vascular blood flow are often interpreted as evidence of biological control. If the flow in a vessel increases, it is assumed that the diameter of the vessel responds in order to auto-regulate the flow. This vasomotion is assumed to be the cause for oscillations in the micro-circulation~\cite{rodgers}. While the importance of vasomotion cannot be denied, there is significant evidence that fluctuations in cell distributions in micro-vascular networks can be due to inherent instabilities~\cite{Kiani:1994aa,Carr:2000aa}.
In 2007 Geddes {\it et al.}
 demonstrated that two important rheological effects are necessary for the existence of multiple equilibria and
 spontaneous oscillations in  microvascular networks ---the F\aa hr\ae us-Lindqvist effect, which governs viscosity of
  blood flow in a single vessel, and the plasma skimming effect,
  which describes the separation of red blood cells at diverging nodes~\cite{Geddes:2007}.

The rheology of blood has been well studied, with the first comprehensive measurements conducted by
  F\aa hr\ae us and Lindqvist in 1931 \cite{Fahraeus:1931aa}.
While blood rheology has many complicated details,
for the existence of multiple equilibria
 the only required ingredient  is that  viscosity is a non-linear function of
red blood cell concentration (hematocrit) \cite{Geddes:2007}.
Such  concentration dependent viscosity relationships are  common to  many systems involving multiple fluids.
In a recent experiment,  we (JBG and BDS) demonstrated that bistability
 exists in a simple network involving two miscible and well mixed
Newtonian fluids of different viscosities~\cite{geddes2010}. These experiments used
 sucrose solution and water, demonstrating
that for  existence of multiple equilibria in the pressure-flow curves, there is no need for  complex  rheology.

Plasma skimming is another source of heterogeneity in microvascular networks   \cite{Geddes:2007}.
Krogh introduced the term plasma skimming in 1921 in order to explain the disproportionate distribution of red blood cells observed at single vessel bifurcations {\it in vivo}~\cite{Krogh:1921aa}. In the absence of plasma skimming, the hematocrit in two downstream vessels of a bifurcation
 would equal that of the feed vessel. Numerous authors
have demonstrated plasma skimming {\it in vitro} and {\it in vivo} and
many attempts to measure the plasma skimming function have been made
\cite{Bugliarello:1964aa,Chien:1985aa,Dellimore:1983aa,Fenton:1985aa,Klitzman:1982aa,Pries:1989aa}.
These measurements have led to
empirical relationships which are a critical component  of the theory and simulation of microvascular flow.

Plasma skimming (or more generally phase separation) exists in numerous other fluid systems. In two fluid systems, it is commonly observed that the phase fraction after a diverging bifurcation is different than that in the feed. Probably the most widely studied example of such phase maldistribution is gas-liquid two phase flow. This system has important technological applications in power and process industries and oil production. In many process applications phase maldistribution can have detrimental consequences for downstream equipment \cite{lahey1986},
while in some cases the phenomenon is exploited to build simple phase separators \cite{Azzopardi1993}. Extensive experimental work on gas liquid flow has been conducted over the past 50 years and these studies have been extensively reviewed
\cite{Azzopardi1999,Azzopardi1994,lahey1986}. All this work has shown that significant separation can occur which is a function of the inlet volume fractions, the geometry of the bifurcation, the fluid properties, and the two-phase flow regime. Simple models can capture some of the experimental features of this system \cite{shoham1987}.

In applications for the  process and petroleum industry, phase separation in liquid-liquid flows are  less well-studied though several recent papers  have emerged \cite{yang2006,yang2007,wang2008}. These studies have typically been conducted with immiscible fluids and at high Reynolds number where the flows are turbulent and often the state of the incoming two-phase flow to the bifurcation is critical to the phase separation. Similar phase separation  effects occur in systems with liquid-vapor flows with evaporation or condensation. The impact of phase maldistribution in two-phase flow has been  shown to impact network flows in  refrigeration systems \cite{hongliang2009} and solar power systems \cite{minzer2006}.

In this work we continue a systematic theoretical and experimental investigation of heterogeneity in simple network flows involving ordinary fluids~\cite{geddes2010}. Our aim is to begin to understand the underlying mechanisms that govern the phase distribution within networks involving fluids with more than one constituent.
We focus on laminar viscous flow of two miscible fluids with different viscosity and density. The fluids stratify in the system due to gravity and remain as nearly distinct phases with some mixing  occurring only by molecular diffusion. This fluid system has the feature that it is easily controlled and modeled, yet has the key ingredients for complex network flows.

In this paper we first use experiments and  3D Navier-Stokes simulations to explore how the two fluids  distribute at a single T-junction.
While such phase separation functions have been extensively measured for gas-liquid flows, to the best of our knowledge they have not been measured in this miscible laminar two-fluid flow.
We measure the phase separation at a single junction, find excellent agreement with 3D Navier-Stokes  simulations, and propose a simple parametric phase separation function. Once the behavior of a single junction is characterized,
 we construct a simple network model. We  find that the phase separation which occurs at a T-junction can lead to  multiple stable equilibria in even the simplest of networks.  Our experiments confirm these predictions.

\section{Phase separation functions}
The first step toward predicting the distribution of phases in a network is to understand the phase separation at a single bifurcation. In microvascular flow, these ``plasma skimming'' functions have received much attention and are a crucial component of network models. In gas-liquid flow, these phase separation functions have also been extensively measured. In order to  make predictions on networks with our fluid system, these phase separation functions must first be measured. Our focus is on laminar stratified flow of Newtonian fluids with different viscosity and density. To the best of our knowledge these phase separation functions have not been measured for this system.  The functions are complicated because they depend upon many parameters in the system. In the first part of this paper, we seek to explore the phase separation at a single T-junction both experimentally and computationally.

\subsection{Experimental system}

\begin{figure}
\begin{center}
\includegraphics[width=4in]{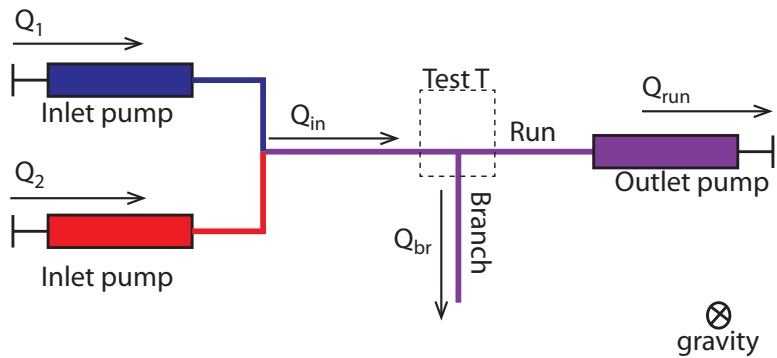}
\end{center}
\caption{Top view schematic of the experimental setup. Two syringe pumps push the two fluids at a controlled rate. Water is in inlet pump 1 and the viscous sucrose
solution is in inlet pump 2. The volume fraction (of fluid 2) into the test T-junction
is given as $\Phi_\mathrm{in}=Q_2/(Q_1+Q_2)$. Fluid is collected from the branch and the volume fraction in the outlet is measured. The outlet pump can be switched to the branch and the experiment repeated.}
\label{fig:schematic}
\end{figure}

A top view schematic of our experimental setup is shown in Figure \ref{fig:schematic}; in this figure gravity points into the page. Two syringe pumps (New Era Pump Systems NE-300) supply our source fluids at a controlled and steady flow rate. Inlet pump 1 contains water (which will be denoted as fluid 1) and inlet  pump 2 contains a controlled aqueous sucrose solution (fluid 2). The mass fraction of sucrose (Sigma Aldrich) in fluid 2 is precisely measured with an analytical balance when the solution is prepared. The viscosity and density of the inlet sucrose solution  are taken from the CRC Handbook from the known mass fraction \cite{RLide2007}. Circular tubing (1.6 mm inner diameter) from the two inlet  pumps meet at the inlet junction where the density difference of the two fluids is sufficient to create a strongly stratified flow.
The dense sucrose solution is observed to sit on the lower half of the tube and the water sits on top. Food coloring is added to both fluids for basic visualization.
The inlet tube then approaches the test T-junction which is held level with respect to gravity (see Figure \ref{fig:schematic}). In the schematic we adopt the nomenclature for the two outlets as ``branch'' and ``run'', terms we will refer to throughout the paper.
The flow rate of one of the outlets is controlled with the outlet syringe pump set in withdraw mode (Harvard Apparatus, ml pump module). The other outlet is held at atmospheric pressure and its flow rate is known from the difference of the pump flow rates. The fluid leaving the system from the open outlet is  collected. The collected
fluid can then  be measured with a Brixometer  (Ataga, PAL-1) to determine the total amount of sucrose in the outlet fluid. The  volume fraction of fluid 2 in the outlet, $\Phi$, can then be determined from this measurement and knowledge of fluid contents of the inlet pumps. The volume fraction  is defined as $\Phi=V_2/(V_1 + V_2)$, where $V$ is the volume and the subscript refers to which fluid.

The test T-junction has a circular cross section of 1.1 mm inner diameter. The entrance and exits to the  T junction are long relative to this diameter; connecting lengths are approximately 100 mm. Flow rates for the incoming fluid are varied through the experiments, but typical rates are on the order of 1 ml/min which corresponds to fluid velocities in the T-junctions of 17 mm/s. The Reynolds number for  water flow at these velocities is $\mathrm{Re}=19$. The Reynolds number of  the sucrose solution of higher viscosity is reduced. The syringe pumps use glass syringes  with a capacity  of either 50 to 20 ml. All pumps were periodically tested and  calibrated for their ability to deliver a steady and correct flow rate.

In a typical experimental run, we wish to measure the volume fraction of fluid 2  in the two outlets as a function of the outlet flow fraction. The outlet flow fraction is defined as the flow in the branch divided by the total inlet flow; $Q_\mathrm{br}/Q_\mathrm{in}$. We set the inlet pumps to a constant flow rate and fixed ratio to control the inlet volume fraction, $\Phi_\mathrm{in}=Q_2/Q_\mathrm{in}$, and total flow rate $Q_\mathrm{in}$. We then vary the outlet withdrawal flow rate of the pump attached to the run. After allowing sufficient time for the system to reach steady state, we collect the fluid from the branch, typically collecting at least 2 ml of fluid. The outlet solution is then well mixed in the collecting container and two successive samples are pipetted and measured with the Brixometer; all readings of the same collection sample fell within the stated error of the device (0.2 \% error in mass fraction). We then change the flow rate of the outlet pump to collect data over a range of outlet flow fractions $Q_\mathrm{br}/Q_\mathrm{in}$. After each change in the flow rate, we let the system come to equilibrium by waiting several flow through times for the network.

With the outlet pump placed on the run, once $Q_\mathrm{br}/Q_\mathrm{in}$ becomes less than about 0.1 we cannot collect a sufficient amount of fluid from the branch before the input syringes empty, thus we end the experiment. We move the pump from  the run to the branch and repeat the procedure, collecting and measuring fluid from the run while controlling the flow in the branch. In this configuration we can  acquire data when the flow in the branch is small, but have difficulty obtaining data when $Q_\mathrm{br}/Q_\mathrm{in}>0.9$ since too little fluid accumulates from  the run before the inlet syringes empty.

While we could collect the contents of the branch (run)  and determine the contents of the run (branch) from conservation, we avoid this approach. We measure the contents of the branch and the run in independent experiments and then confirm that the inferred values from conservation are in agreement with the measured values. It should be noted that the inferred values of the branch contents (when the run is measured) are unreliable when $Q_\mathrm{br}/Q_\mathrm{in}>0.9$. The sensitivity in the calculation is such that a small error in the measured value leads to a large error in the inferred value, thus the measured values are to be trusted over the inferred. We repeat each measured data point  three times. Each of these three measurements is a unique experiment; new inlet fluids are mixed, all syringes are cleaned, and a new identical network is constructed from new materials and components.

\subsection{Simulation}
We conducted simulations of the system using a commercial 3D finite element code, Comsol Multiphysics. The simulations solve the steady state,  incompressible Navier-Stokes equations for conservation of momentum,
\begin{equation}
\rho \mathbf{u} \cdot \nabla \mathbf{u} = -\nabla P + \rho \mathbf{g} + \nabla \cdot \left(\mu \left(\nabla \mathbf{u} + \nabla \mathbf{u}^T \right)\right),
\end{equation}
and mass
\begin{equation}\
\nabla \cdot \mathbf{u}=0.
\end{equation}
We solve the convection-diffusion of a dilute species which represents the relative concentration of sucrose,
\begin{equation}
\mathbf{u} \cdot \nabla \mathbf{c} = \nabla \cdot \left( D \nabla \mathbf{c} \right).
\end{equation}
 Here, $\mathbf{u}$ is the velocity vector, $\rho$ is the density, $\mu$ is the viscosity, $c$ is the concentration, and $D$ is the diffusivity. The concentration equation couples back to the momentum equation through the dependence of the density and viscosity on  concentration. We approximate the fluid viscosity  as,
\begin{equation}
\frac{\mu}{\mu_1} =  \left( \frac{\mu_2}{\mu_1} \right)^{(c^2+c)/2} ,
\end{equation}
where $\mu_1$ is the viscosity of water (fluid 1) and $\mu_2$ is the viscosity of the inlet sucrose solution (fluid 2). The power of $(c^2+c)/2$ was empirically fit to match the experimental data with reasonable accuracy \cite{RLide2007}. Our definition of concentration, $c$, is normalized by the concentration in fluid 2, thus the concentration in our model varies between 0 and 1.  The density is assumed to depend linearly on concentration, $\rho = \rho_1 + (\rho_2 - \rho_1) c$, where $\rho_1$ and $\rho_2$ are the densities of fluid 1 and  2 respectively. The diffusivity is assumed to vary inversely with the viscosity as $D = D_1 \mu_1/\mu$.
Textbook values for the diffusivity of sucrose in water at room temperature are around $D_1=5 \times 10^{-10}$ $\mathrm{m^2/s}$.

\begin{figure}
\begin{center}
a) \includegraphics[width=2.7in]{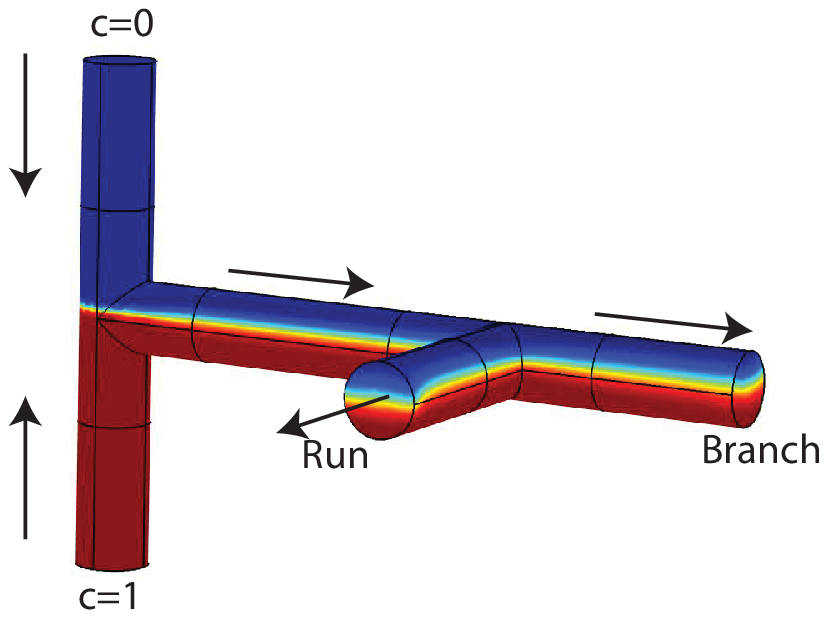}
b) \includegraphics[width=3in]{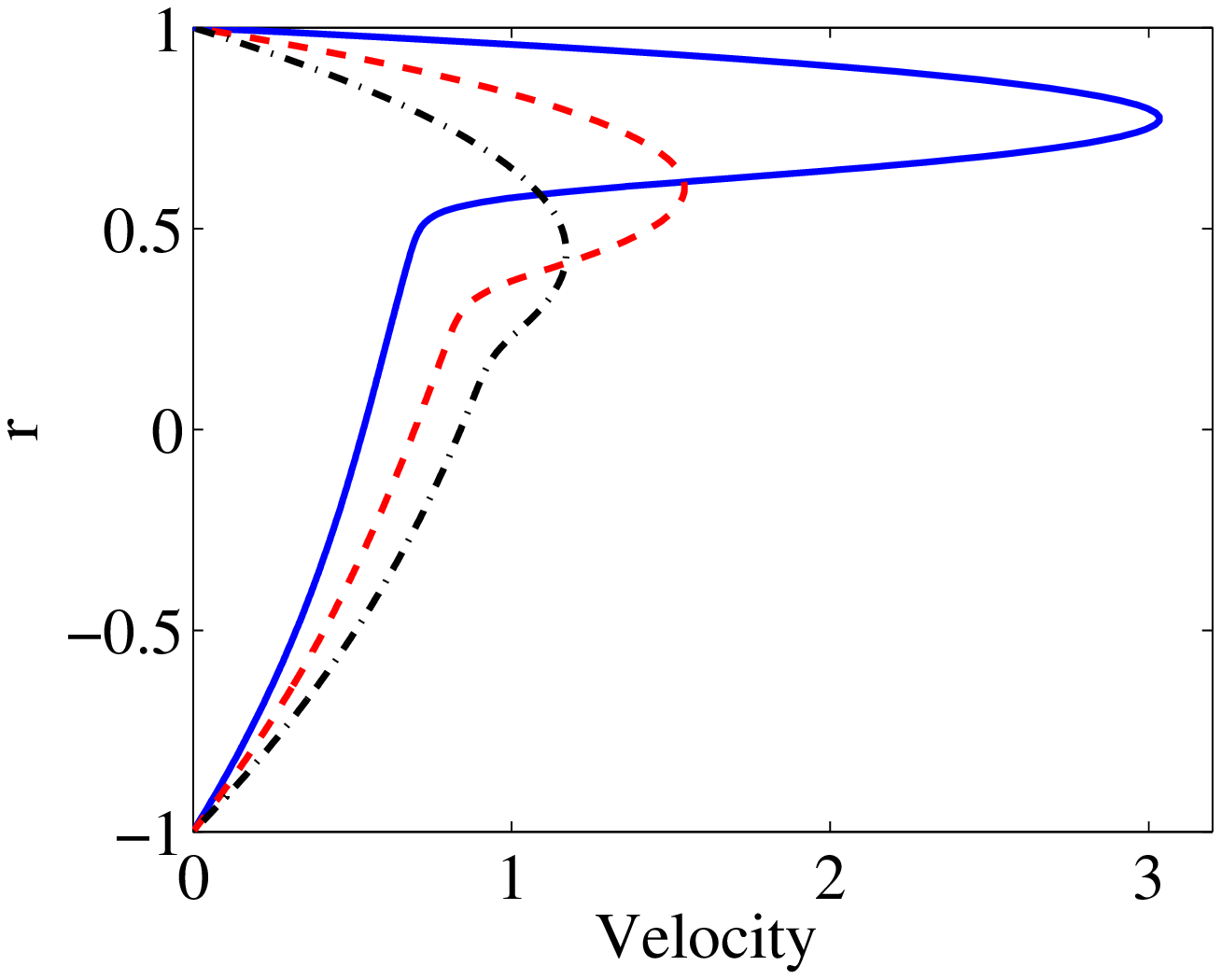}\\
\end{center}
\caption{a) Simulation geometry. The color represents the concentration field. The two fluids are brought in vertically with respect to gravity so that the flow will be pre-stratified in the inlet.  b) Sample velocity profiles for stratified flow as a function of radial distance from the center of the tube. Profiles are shown for viscosity ratios of 5 (dash-dot black curve), 15 (dashed red), and 100 (solid blue).  For each case the  inlet volume fraction is  $\Phi_\mathrm{in}=0.5$. Notice that the location of the interface between the two fluids moves toward the wall as the viscosity ratio increases.  The radial coordinate in the plot is taken along a vertical line through the center of the tube in the direction of gravity. For a viscosity ratio of 1 we would have the classic parabolic velocity profile. }
\label{fig:simulation}
\end{figure}

The simulation domain, shown in Figure \ref{fig:simulation}a, consists of two inlet tubes which are aligned with respect to gravity. The inlet concentration is 0 in the upper inlet (pump 1)  and 1 in the lower inlet (pump 2); both inlet flow rates are controlled. The two fluids meet and merge in a single tube, approach  the test T junction and then proceed to the outlets. The stratified flow in the approach to the test T is enforced by aligning the inlet tubes  with gravity. In the experiment the entrance is configured as shown in Figure 1, however, the long entrance length ensures fully stratified flow at the T-junction. In the computation it is not practical to have a long entrance length before the T-junction. The outlet flow rate is controlled on the branch through the boundary condition  while the outlet at the run is allowed to freely flow out to a reservoir. The simulation was run on progressively finer grids to ensure adequate convergence in the solution.

We make the equations dimensionless by using the pipe diameter, $d$,  as the length scale,  and the total input flow rate divided by the area of the pipe as the velocity scale, $U_0$. Under this scaling the equations become,
\begin{equation}
(1 + \beta c)  \mathbf{u} \cdot \nabla \mathbf{u} = -\nabla P + \mathrm{Fr}  \beta c  \mathbf{\hat{z}} + \frac{1}{\mathrm{Re}} \nabla \cdot \left( {\tilde{\mu}} ^{(c^2+c)/2} ~\left( \nabla \mathbf{u}  +\nabla \mathbf{u}^T \right)\right),
\end{equation}
\begin{equation}\
\nabla \cdot \mathbf{u}=0,
\end{equation}
\begin{equation}
\mathbf{u} \cdot \nabla \mathbf{c} = \frac{1}{\mathrm{Re}~\mathrm{Sc}} \nabla \cdot \left( {\tilde{\mu}} ^{-(c^2+c)/2} \nabla \mathbf{c} \right).
\end{equation}
There are six dimensionless parameters from the flow model: the Reynolds number $\mathrm{Re}=\frac{\rho_1 U_0 d}{\mu_1}$; the Froude number $ \mathrm{Fr}=g d/U_0^2$;  the viscosity ratio $\tilde{\mu} = \mu_2/\mu_1$; the inlet volume fraction $\Phi_{\mathrm{in}}$; the Schmidt number $\mathrm{Sc} = \mu_1/ \rho_1 D_1$; and the density difference ratio $\beta =  (\rho_2 - \rho_1)/\rho_1$. The geometry  provides the ratio of the diameter to the length of the entrance
tube $d/L$ as an additional parameter.
The Schmidt number is a material constant with a typical value of $\mathrm{Sc}=2000$ and is thus not a free parameter that we can easily control.
As with other flow problems where gravity is important, the Reynolds number and Froude number are related in a way that they  cannot be independently varied.

The entrance  length $L$ only plays a role in determining
how much diffusive mixing occurs in the tube leading up to the T-junction.
The time scale for diffusive mixing in the tubes is approximately $d^2/D \sim 2400$ seconds. The flow through time for the entrance tube is $L/U \sim 6$ seconds (the exact value depending on the flow rate). Thus for all the cases presented in this paper the two phases have little time to diffusively
mix before entering the T-junction.

When two fluids of different viscosity flow inside a tube, it is important to realize that
the volume taken up inside the tube is not equal to the ratio of the flow rates. The low viscosity water is squeezed to the top of the tube where it has a higher velocity than the viscous sucrose solution.
Sample velocity profiles in the tubes are shown in Figure \ref{fig:simulation}b.
Further, we must also ask whether these laminar velocity profiles are stable.
The stability of stratified flow of two fluids different viscosity has been widely studied due to its industrial relevance. It has been demonstrated that viscosity stratified Poiseuille flows  can be unstable even in the low Reynolds number and Stokes flow regime \cite{Yih1967,Talon2011}. In our work, we have strong buoyancy effects which keep the flow stably stratified. In all experiments we observe that the interface between the two fluids remains sharp and free of any obvious instability.

\subsection{Results}

\begin{figure}
a) \includegraphics[width=3.in]{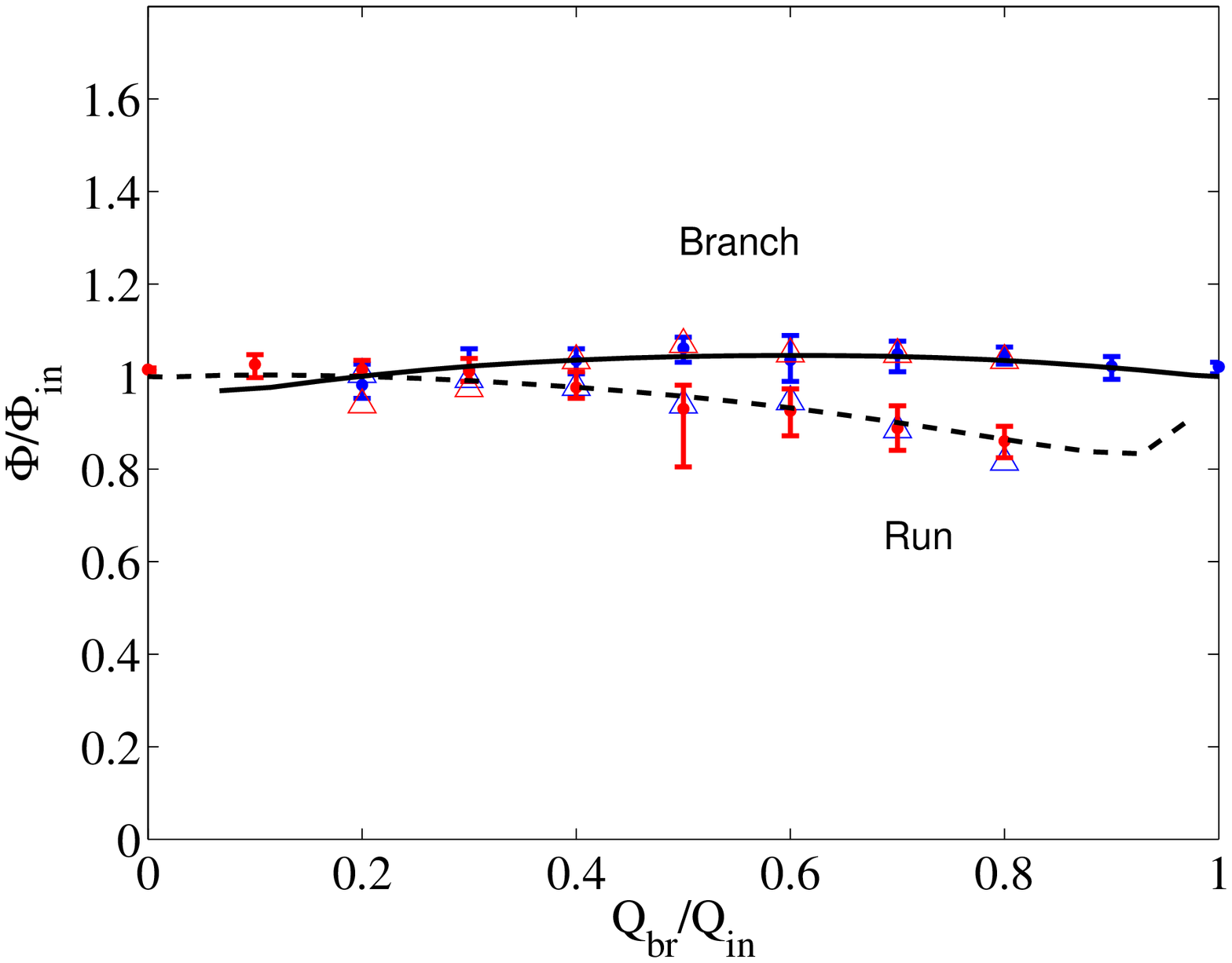}
b) \includegraphics[width=3.in]{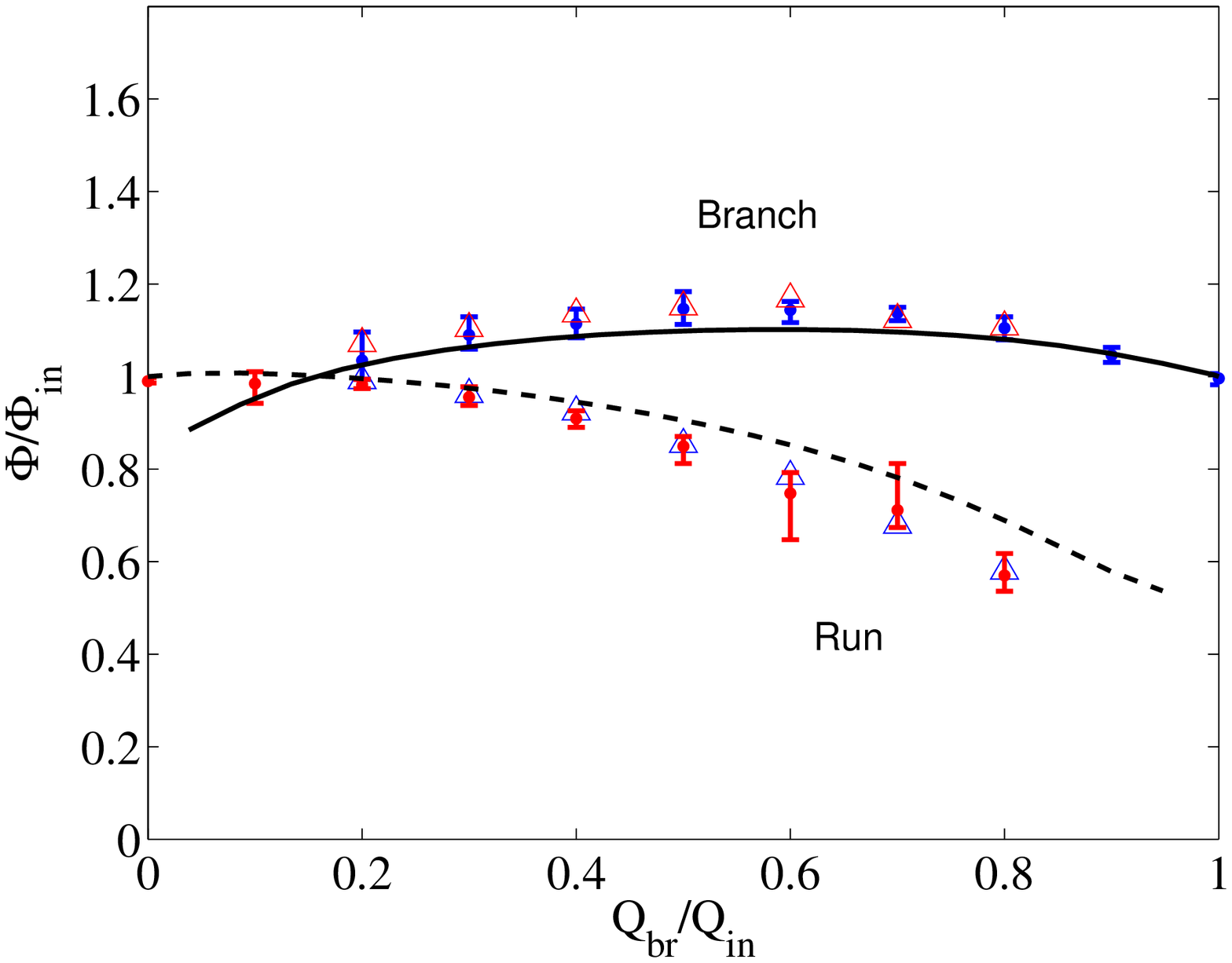}\\
c) \includegraphics[width=3.in]{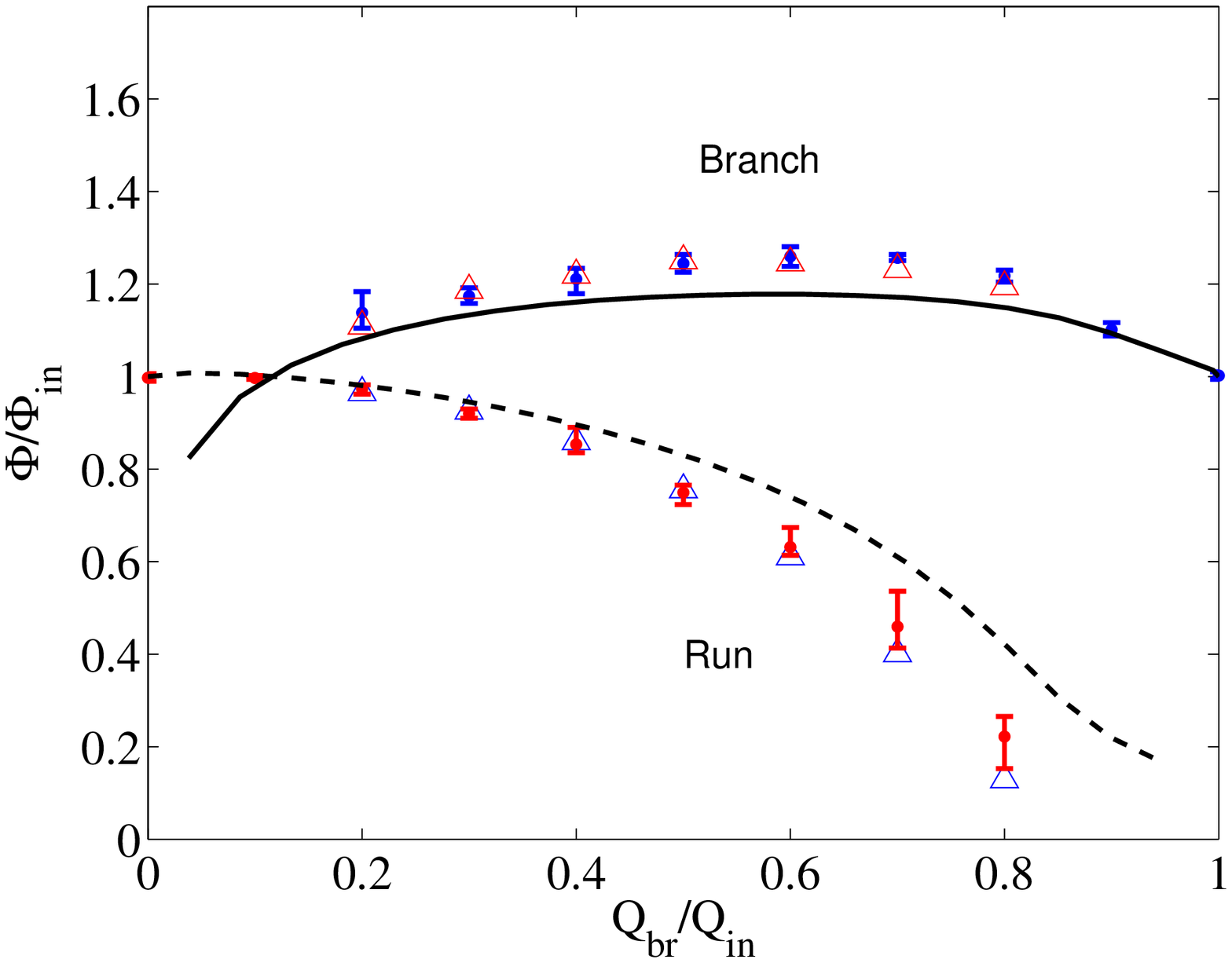}
d) \includegraphics[width=3.in]{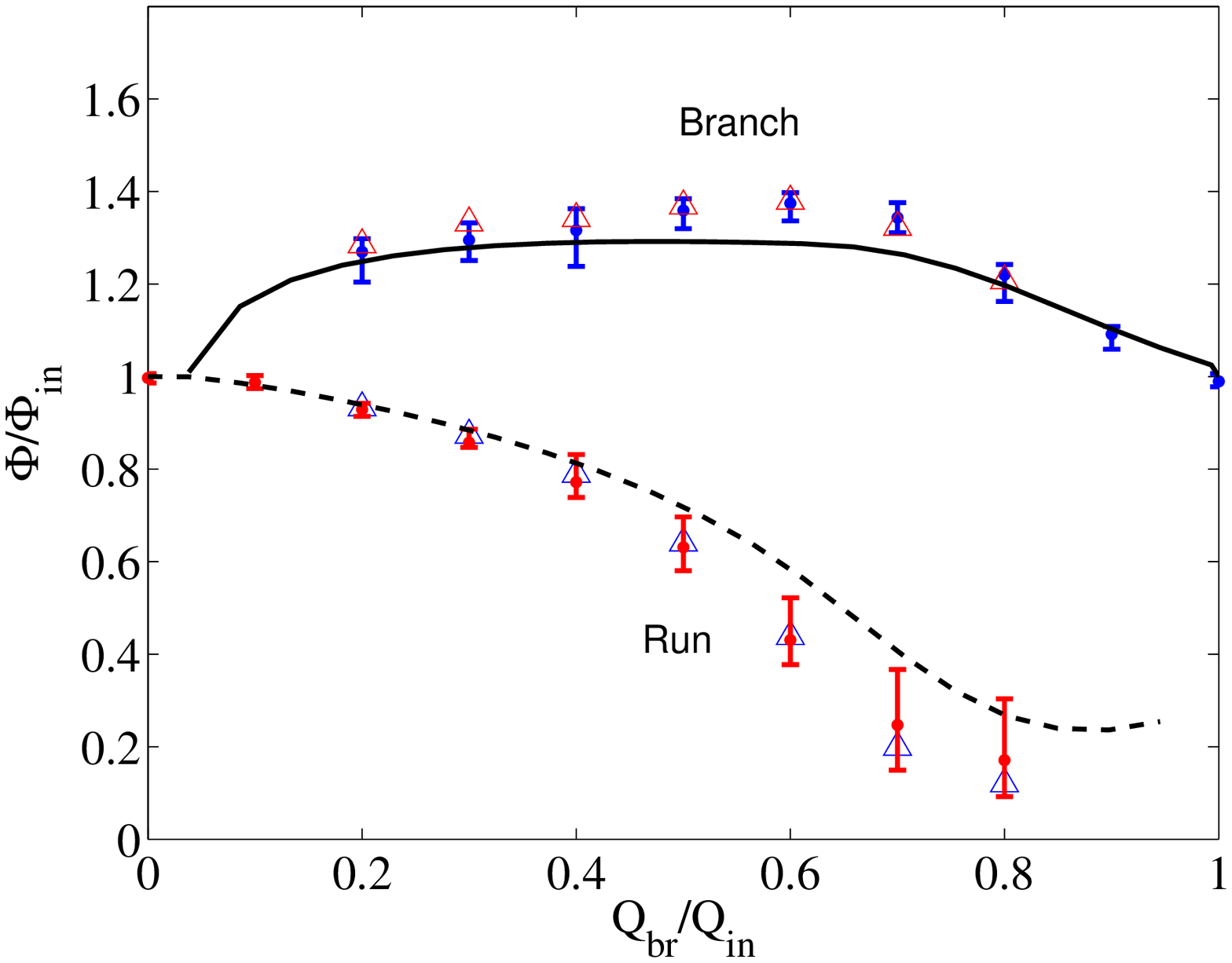}
\caption{Comparison of experiment and simulation for the phase splitting at a T junction as a function of the overall flow rate. Each inlet pump has a flow rate of  a)  0.5 ml/min, b) 1 ml/min, c) 2 ml/min and d) 5 ml/min. Each data point represents the mean of six measurements (2 independent measurements on 3 independent but identical networks). The error bars represent the maximum and minimum values measured. The composition of the branch and the run are measured separately, though this measurement is redundant. The triangles represent the mean value in the branch (run) as calculated from the measurements in the run (branch)---the data are consistent. The red data points are data measured  in the run and the blue points  are measured in the branch. The solid line is the finite element simulation from the branch and the dashed line is the simulation from the run. The sucrose solution coming from pump 2 is a 50\% mass fraction solution which has a viscosity of $\tilde{\mu}=15$.  The Reynolds number in the T-junction (based on the water viscosity) for each case is approximately  a) 19, b) 38, c) 76, and  d) 190. The Froude number in each case is a) 35 b) 8.7 c) 2.2 d) 0.35. The other dimensionless parameters are $\mathrm{Sc}=2000$ and $\beta=0.23$. }
\label{fig:flow}
\end{figure}

In Figure \ref{fig:flow} we show a comparison of the experimental measurements and 3D simulations at four different total inlet flow rates. In each sub-figure we plot the volume fraction of the fluid in the branch and in the run, divided by the inlet volume fraction as we vary the flow ratio, $Q_\mathrm{br}/Q_\mathrm{in}$. For all cases the inlet volume fraction is held fixed at $\Phi_\mathrm{in}=0.5$ and $\tilde{\mu}=15$, but the overall inlet flow rate is varied. If there were no phase separation effect, then each plot would show unity for all values of the flow fraction. However, we see significant phase separation which depends strongly on the overall flow rate or inlet Reynolds number. Each point represents the average of three independent trials and the error bars represent the spread in these trails. The triangles represent the inferred values from the measurements of the other outlet. Across the range of experiments the measured and inferred values are in excellent agreement, giving us high confidence in the experimental procedure.

In general, we find good agreement between the simulation and the experiment. The agreement is best at low flow rate while at higher flow the simulation under-predicts the experimental data. However, the predicted trend is in excellent agreement for  a non-linear flow problem. Since the amount of separation depends sensitively on the total flow rate, it is clear that inertial non-linearity in the Navier-Stokes equations is the origin of the phase separation.
 In dimensionless terms, each  experiment shown in Figure \ref{fig:flow} has a different Reynolds number and  Froude number.

 The mechanism for the splitting is an inertial, or Reynolds number, effect---as the fluid mixture approaches the T, the fast moving water tends to overshoot the 90 degree bend and continues onward along the run. The run always contains more water than the inlet fluid.
  We use simulation to make sure that the phase separation's dependence is not
   due to changes in the Froude number. We ran simulations with gravity turned off and get very similar results as those
   presented in Figure \ref{fig:flow}. The results are relatively insensitive to the Froude number.

The data near the end points (when $Q_\mathrm{br}/Q_\mathrm{in}$ is close to zero or one) are difficult to resolve both for the simulation and the experiment. For example,  when the flow in the branch approaches zero, the volume fraction in the run must approach that of the inlet. The branch can have essentially any volume fraction and leave the run unchanged. In the experiment, we cannot collect enough fluid  at these conditions before  the syringe pump empties, thus the contents cannot be measured directly.  We also cannot calculate the contents of the branch from the contents of the run, unless we know the run contents with extreme precision---the error gets amplified in this regime. Computationally, the flow and flux of concentration  in the branch are both very close to zero near this point. Therefore, the ratio of the two is susceptible to a small amount of error in the overall numerical solution. With the current arrangement, we cannot confidently say what the concentration in the branch (run) is as the branch flow rate approaches zero (one). The simulation is well behaved in the range $0.05 < Q_\mathrm{br}/Q_\mathrm{in} < 0.95$. Beyond this range, while the numerical solution returns a well converged solution, the calculation of the volume fraction in the low flow outlet is not well behaved - exactly as in the experiments.

\begin{figure}
a) \includegraphics[width=3.in]{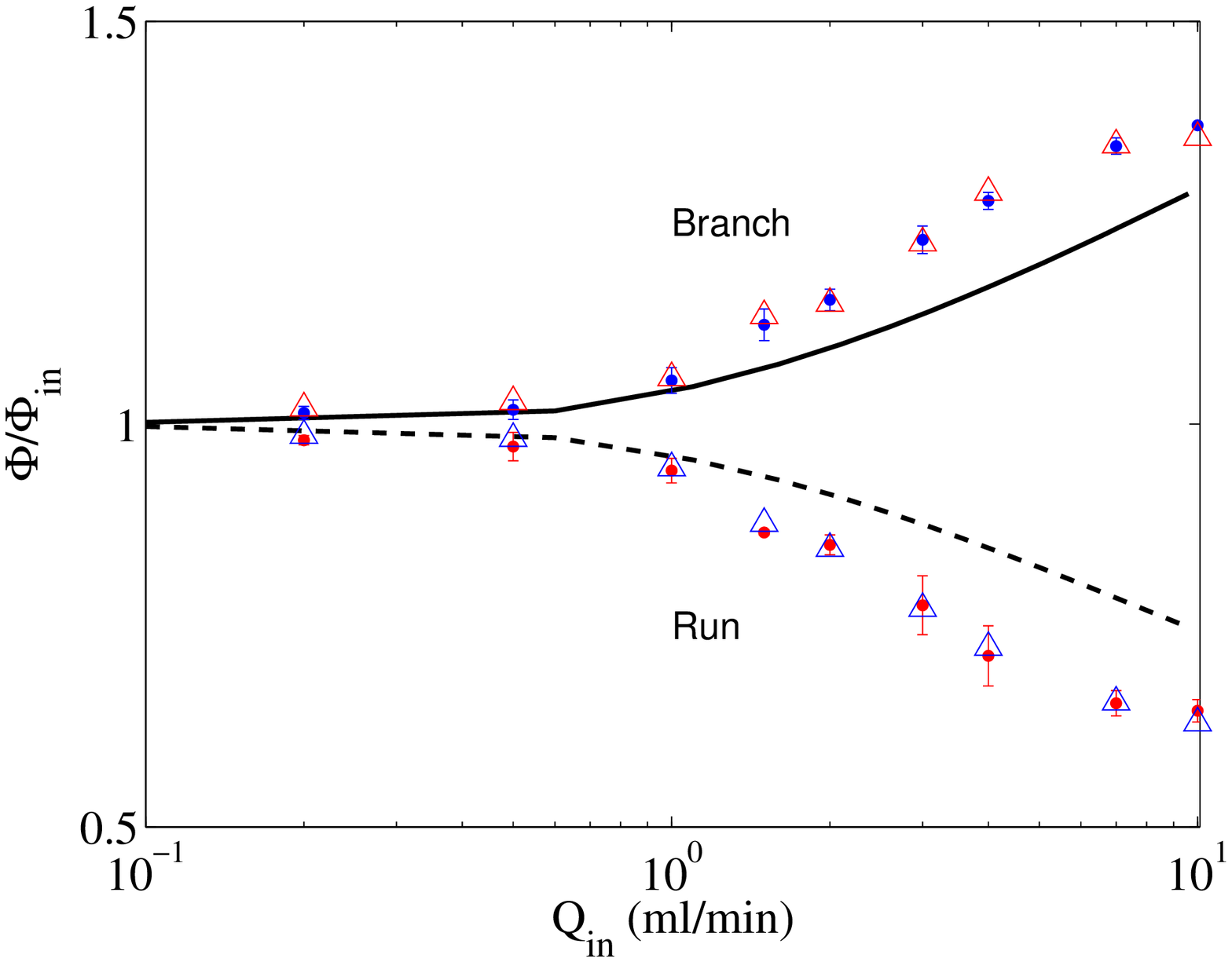}
b)  \includegraphics[width=3.in]{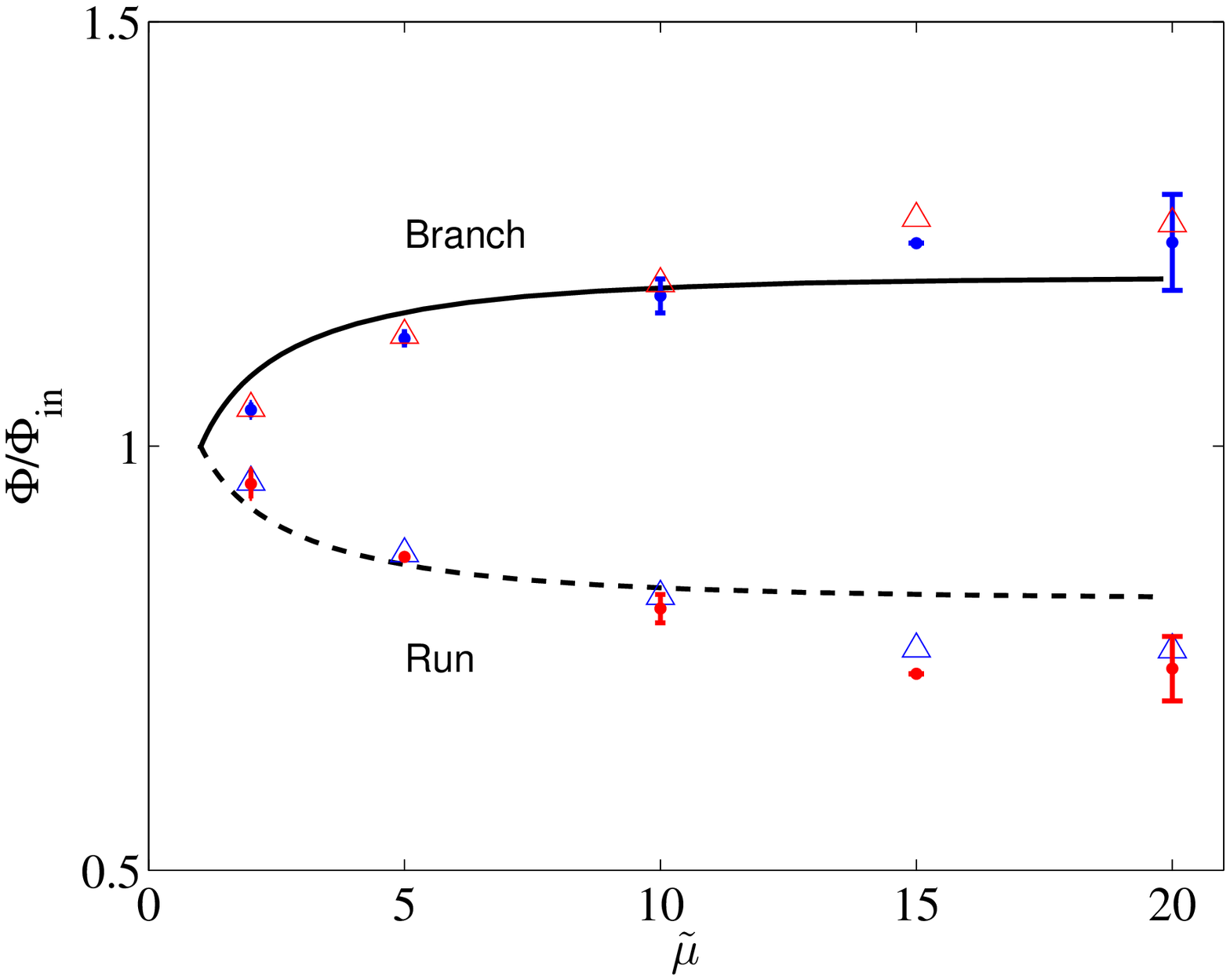}\\
\caption{Comparison of experiment and simulation. Here the two outlet flows are fixed to always be equal; $Q_\mathrm{br}/Q_\mathrm{in}=0.5$.
In a)  we only vary the total flow rate holding the viscosity ratio at $\tilde{\mu}=15$.
In b) we hold the total flow constant at 4 ml/min ($\mathrm{Re}=76~\mathrm{Fr}=2.2$) and vary the viscosity ratio.
In all cases $\mathrm{Sc}=2000$ and $\beta=0.23$.
The symbols are the same as in Figure \ref{fig:flow}.   }
\label{fig:flow_visc}
\end{figure}

In addition to varying the inlet flow rate, we also explored  how the phase separation depends upon other parameters in the system. In Figure  \ref{fig:flow_visc} we hold the outlet flow fraction fixed at 0.5 (i.e. $Q_\mathrm{br}=Q_\mathrm{run})$
and vary both the total flow and the viscosity contrast. The inlet volume fraction is $\Phi_\mathrm{in}=0.5$.
In Figure \ref{fig:flow_visc}a we vary the overall flow rate while in Figure \ref{fig:flow_visc}b
we vary  the viscosity ratio.
As before we measure the contents of the branch and the run separately, using three separate trials for each data point.
The Comsol simulation tends to under-predict the phase separation, however the trend in both cases is well-captured.

\begin{figure}
a) \includegraphics[width=3.in]{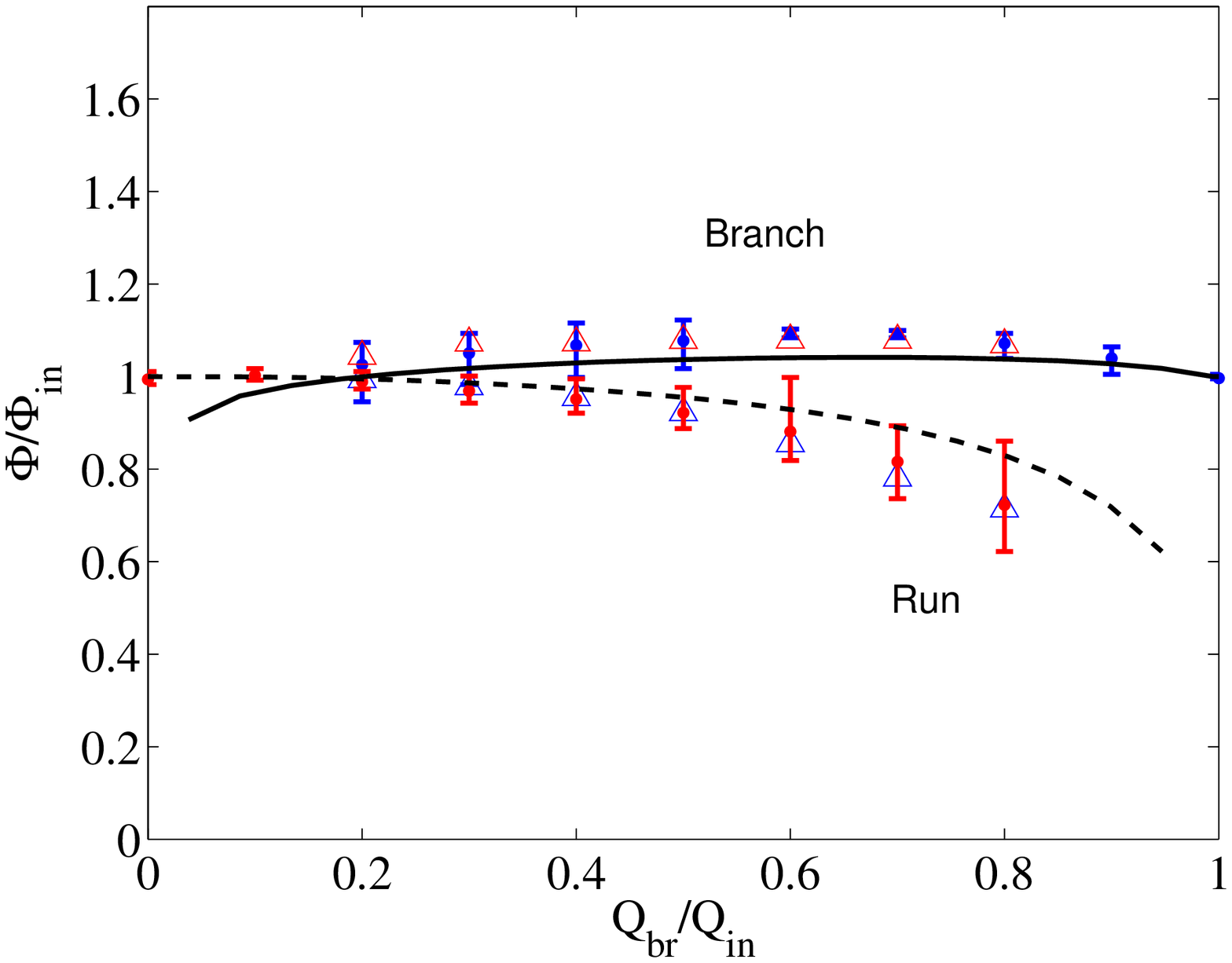}
b) \includegraphics[width=3.in]{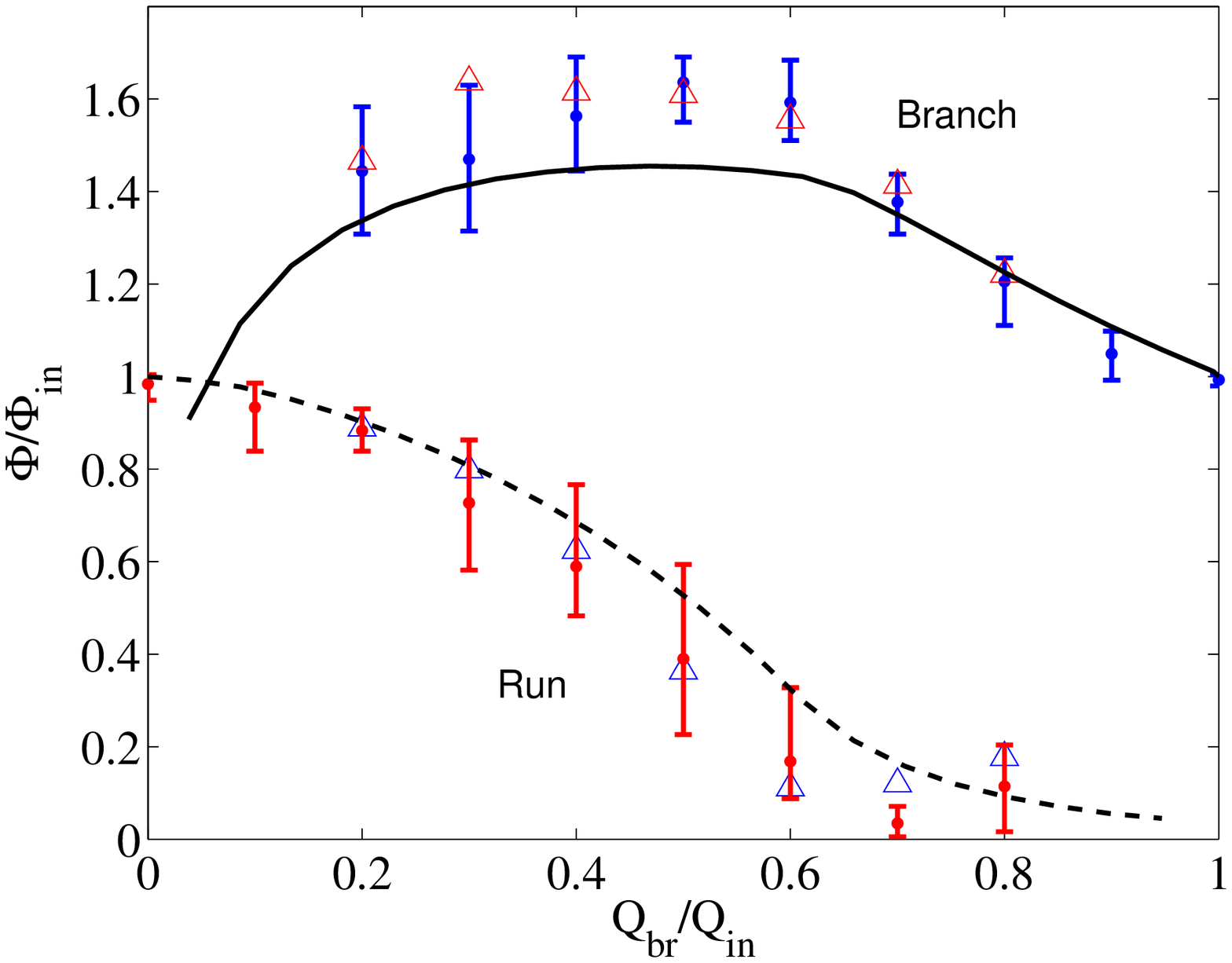}\\
\caption{Comparison of experiment and simulation for the phase splitting at a T junction.
In both experiments the total inlet flow rate is 4 ml/min. In  a)  the sucrose inlet is 2.8 and the water is 1.2  ml/min, $\Phi_{in}=0.7$. In  b) the sucrose is 1.2 and the water is 2.8 ml/min, $\Phi_{in}=0.3$.
The sucrose solution in the inlet pump has a viscosity 15 times that of water.
The other dimensionless parameters are $\mathrm{Re}=76,~\mathrm{Fr}=2.2,~\beta=0.23$ and $\mathrm{Sc}=2000$.
The symbols are the same as in Figure \ref{fig:flow}. }
\label{fig:phi_in}
\end{figure}

In Figure \ref{fig:phi_in} we show the phase separation as we change the inlet volume fraction. In this case we hold the total flow at 4 ml/min and show results for $\Phi_{\mathrm{in}}=0.7$ and $\Phi_{\mathrm{in}}=0.3$. The viscosity ratio between the two fluids is $\tilde{\mu}=15$.
We find that the phase separation depends quite strongly on the incoming volume fraction, and that the separation is quite dramatic when the volume fraction of fluid 2 is low.

In general we see find very good agreement between the experiment and simulations. Typically,  the simulation under-predicts the amount of phase separation, however
all trends seem to be well captured by the finite element simulation. The simulation, therefore, is a useful tool for these flows. We can use the simulation to predict phase separation functions over a much wider range of parameter space,  much more quickly than we can experimentally.  Given the good agreement we find here, we have confidence that predicted separation functions will be  accurate.
To the best of our knowledge, these measurements and simulations represent the first in this geometry with laminar, stratified,  miscible flow.

\begin{figure}
\begin{center}
 \includegraphics[width=3.in]{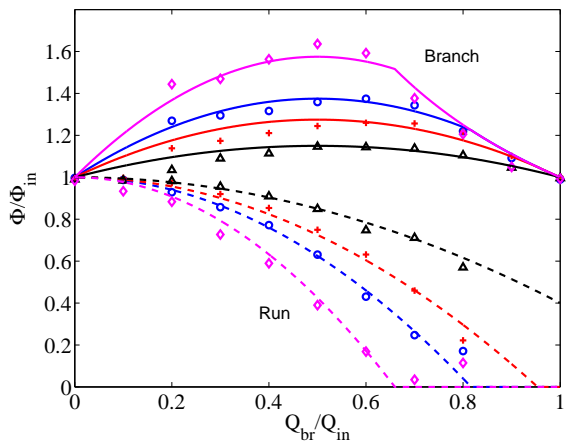}
\end{center}
\caption{ Comparison of experimental data for the phase splitting at a T junction compared to the simple 1 parameter model over
a range of conditions.
The points are experimental data (error bars removed for clarity) and
the lines are from  the fit function - solid lines for the branch and dashed lines for the run.
All experimental data is the same as shown in previous figures.
The conditions shown are;
i) black lines for $\alpha=0.6$ and black triangles for 2 ml/min total flow with  $\Phi_{\mathrm{in}}=0.5$;
ii) red lines for $\alpha=1.1$ and red pluses for 4 ml/min total flow with  $\Phi_{\mathrm{in}}=0.5$;
iii) blue lines for $\alpha=1.5$ and blue circles for 10 ml/min total flow with  $\Phi_{\mathrm{in}}=0.5$;
iv) magenta lines for $\alpha=2.2$ and magenta diamonds for 4 ml/min total flow with  $\Phi_{\mathrm{in}}=0.3$;
In all cases the sucrose solution in the inlet pump has a viscosity 15 times that of water.
}
\label{fig:fitfcn}
\end{figure}

As we will explain in the next section, these measured separation functions are a key ingredient in network models.
The separation function must be applied at each node within the network.
In order to quickly explore ramifications of phase separation for network flows,
we empirically develop a simple one-parameter model function for $\Phi_\mathrm{br}$
and $\Phi_\mathrm{run}$. We assume
\[
\frac{\Phi_\mathrm{br}}{\Phi_\mathrm{in}} =  \left(1 + \alpha \left(1 - \frac{Q_\mathrm{br}}{Q_\mathrm{in}}\right) \frac{Q_\mathrm{br}}{Q_\mathrm{in}} \right),
\]
and that $\Phi_\mathrm{run}$ is found from conservation
\[
\frac{\Phi_\mathrm{run} }{\Phi_\mathrm{in}} =  \left( 1  -  \alpha \left(\frac{Q_\mathrm{br}}{Q_\mathrm{in}} \right)^2 \right).
\]
If the control parameter $\alpha$ is greater than 1 we use a piecewise approximation where
$\Phi_\mathrm{run}=0$ in the regions where the volume fraction would be negative.
If $\Phi_\mathrm{run}=0$, then $\Phi_\mathrm{br}$ is computed from conservation.
The construction of this simple function, while far from precise captures the basic trends of our experimental data. This empirical function allows us to scan parameter space easily in making network predictions. With this simple function, $\alpha$ is related to the strength of the phase separation. An example of the fit function compared to experimental data is shown in Figure \ref{fig:fitfcn}.

We emphasize that this simple function and the control parameter $\alpha$ are empirically defined and only done for convenience in the network simulations.
The parameter $\alpha$ depends  the inlet volume fraction,  the inlet Reynolds number, and the viscosity contrast. The data fit with this function in Figure \ref{fig:fitfcn} is not sufficient in quantity to create a fit for $\alpha$ over all of the parameter space.
We only use the empirical phase separation function to  provide some insight into the network behavior through a simple model.

\section{Network flows}
\begin{figure}
\begin{center}
 \includegraphics[width=4.5in]{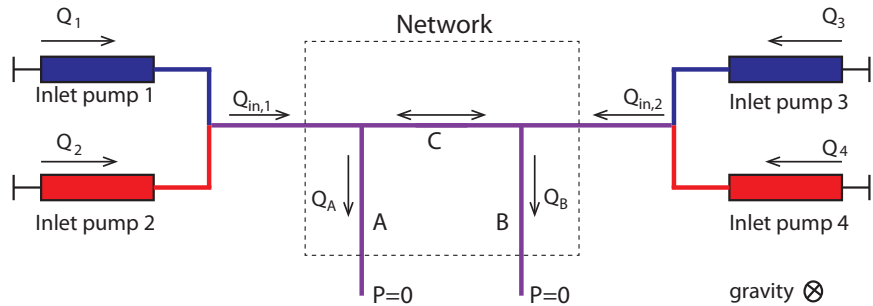}
\end{center}
\caption{
 Top view schematic of the network as constructed in experiment.
 The results are sensitive to whether the connector between the two outlets is the run
(as configured here) or the branch of the T-junction. }
\label{fig:network}
\end{figure}

Now that the phase distribution functions are known for a single junction,
we can begin to use these functions to predict how the phases would distribute within a network.
In this paper we use one of the simplest networks possible, Figure \ref{fig:network},  which we demonstrated in previous work had bistability when the two inlets ($Q_{\mathrm{in},1}$ and $Q_{\mathrm{in},2}$) were different fluids and the viscosity contrast was sufficiently strong~\cite{geddes2010}.
In this work we consider the same network, only the two inlets to the network are the stratified 2-component flow rather than single phase fluids. We find a different type of bistability which originates from the phase distribution function at the single T-junction,  occurring when the two inlets to the network are identical fluid mixtures.
In this setup,  we use four inlet pumps to create a controlled stratified flow as the two inlets to the network.
The inlet volume fractions to the network are defined by $\Phi_{\mathrm{in},1} = Q_2/(Q_1+Q_2)$ and $\Phi_{\mathrm{in},2} = Q_4/(Q_3+Q_4)$. The total flow into the network is $Q_{\mathrm{tot}} = Q_{\mathrm{in},1} + Q_{\mathrm{in},2}$.

\subsection{Analysis}

\begin{figure}
\begin{center}
a) \includegraphics[width=3.in]{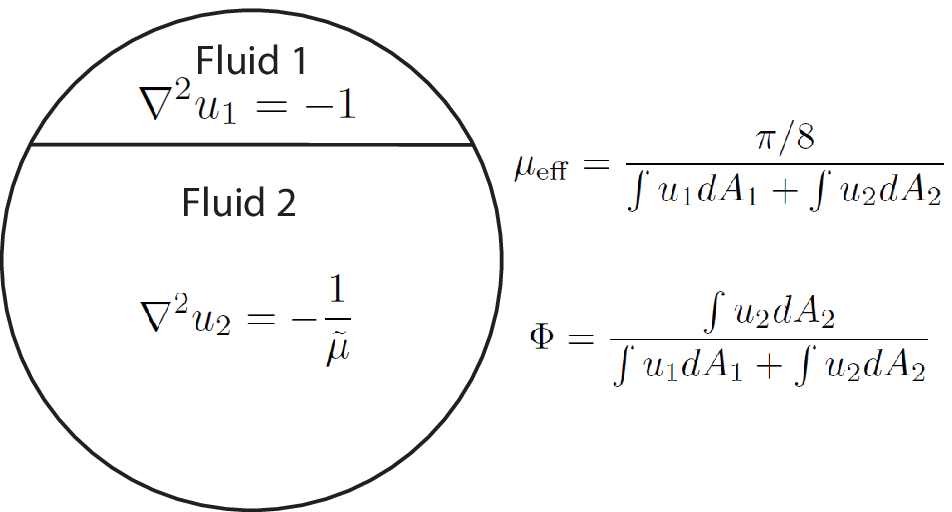}
b) \includegraphics[width=3.in]{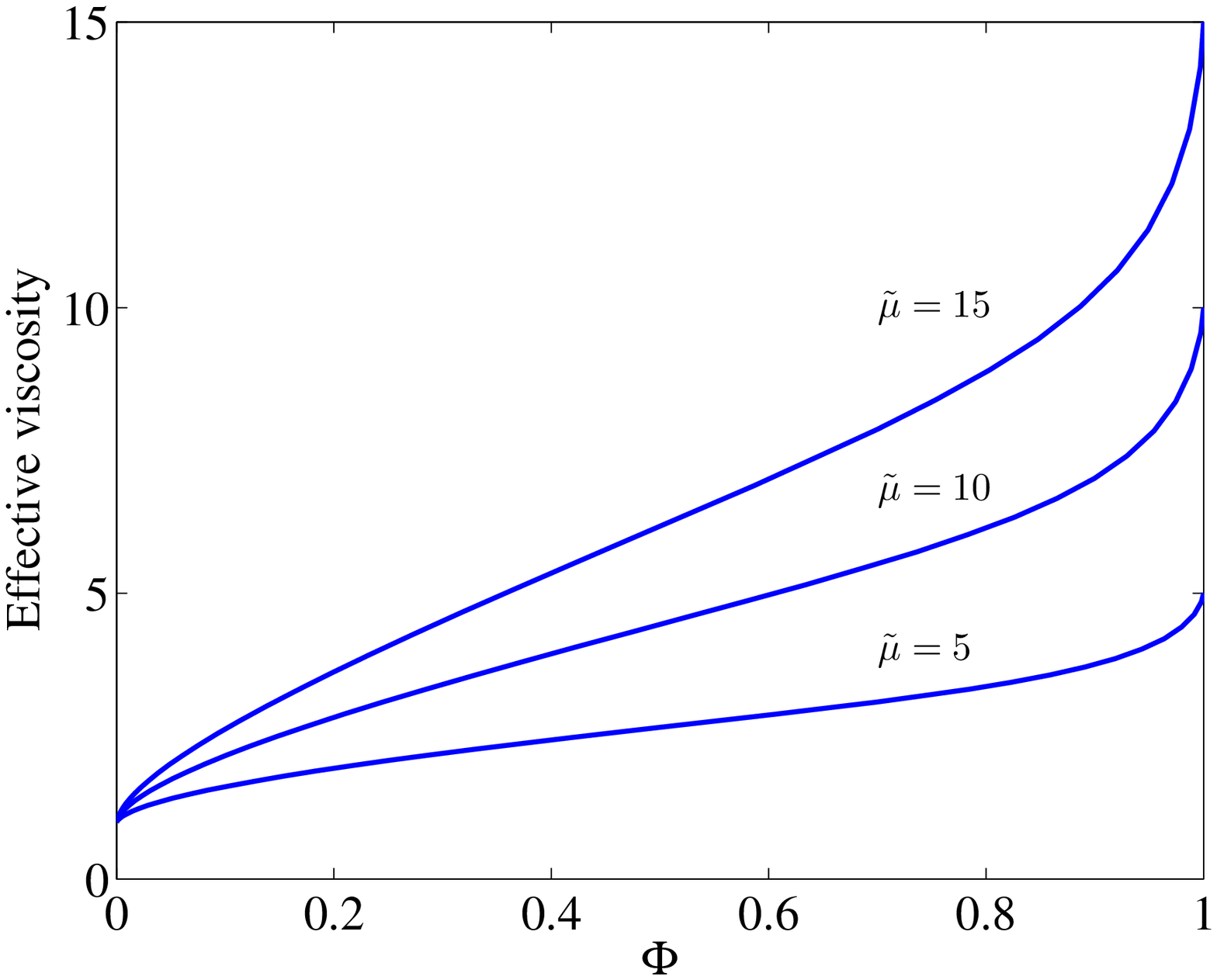}
\end{center}
\caption{a) Schematic and formulation for calculating the effective viscosity of miscible stratified flow in a tube. We show a cross section of the tube
 where $u$ is the axial velocity coming out of the page. The boundary conditions are no-slip at the wall and constant stress at the interface between the two fluids. b) Effective viscosity for miscible stratified laminar flow of two fluids in a tube as a function of volume fraction of the more viscous fluid in the mixture. Sample curves are shown for viscosity contrast of 5, 10, and 15.   The normalized diffusion time measured in units of  $D/d^2$   was 0.0025; the final result is very close to the immiscible case.  }
\label{fig:visc}
\end{figure}

The pressure drop across any length of tube can be described by
\[
\Delta P = Q R
\]
where $R$ is the hydraulic resistance of the tube, $\Delta P$ is the pressure drop and $Q$ is the volumetric flow rate. The resistance for a single phase laminar flow is simply Poiseuille's Law, $R=128 \mu L/\pi d^4$ where $L$ and $d$ are the length and diameter of the tube respectively.
For our two-fluid system the viscosity in Poiseuille's law is replaced by an effective viscosity.
The effective viscosity is calculated from the full velocity profile of the two fluids in contact with each other in the tube, as in Figure \ref{fig:simulation}b.
In the flow regime of our network, the resistance behavior is  similar to what one would calculate for immiscible fluids in fully developed flow \cite{Gemmell:1962p5510}.
The setup for the calculation of the effective viscosity of immiscible fluids
is shown in Figure \ref{fig:visc}a.
Since the flow is fully developed along the axial length of the tube, we solve for the velocity across a cross section of tube assuming a horizontal interface separating the two fluids. Since the velocity only varies across the cross section the Navier-Stokes equations simplify dramatically as illustrated in the schematic.
Once the velocity profile is computed for a fixed interface location, we can easily calculate the overall
hydraulic resistance and the volume fraction of the two flows.

Since our system has miscible fluids, we can solve the same  problem only for an interface which has been smeared by diffusion.
When diffusion exists in the system,  the effective resistance of the flow in the tube depends on
how long the two fluids have been in contact with each other in the tube.
For long and narrow tubes, there would be sufficient time for molecular diffusion
 and the behavior limits to the fully mixed case.
 For short tubes, we would limit to the immiscible result. The relevant dimensionless parameter would compare the time scale for diffusion across the cross section of tube relative to the transit time through the tube; i.e. $u d^2/D L=\mathrm{Re}~ \mathrm{Sc}~\frac{d}{L}$. At the lower Reynolds numbers
 of our experiments
 this parameter is approximately 400, meaning the time scale for  diffusion is 400 times slower than the transport time through the tubes.
 Example curves for the effective viscosity (based on miscible fluids) are shown in Figure \ref{fig:visc}, which fall very close to the immiscible result (not shown).
The calculation of the effective viscosity and the effect of diffusion on the resistance in a single tube was discussed in more detail in our previous work \cite{geddes2010}.

Once the effective viscosity as a function of the volume fraction  is known, we can relate the pressure drop and flow in each of the tubes.
For our network,  the flow is defined as positive for  flowing out  at exits A and B and positive flow in C is defined from left to right.
 With this convention the pressure drops in the three tubes  must satisfy,
\[
\Delta P_A = \Delta P_C + \Delta P_B
\]
Using flow conservation, we know that $Q_A = Q_{\mathrm{in},1} - Q_C$ and $Q_B = Q_{\mathrm{in},2} + Q_C$.
We can  describe the flow in tube C as,
\[
Q_C= \frac{Q_{\mathrm{in},1}  R_A -  Q_{\mathrm{in},2}  R_B }{R_A + R_B+R_C}.
\]
It is important to remember that the resistances of each tube in the above expression depends upon the volume fraction in the tube.

When the flow in C is positive then phase separation occurs at the T-junction on the left and thus this phase separation determines the contents of tubes A and C.
The volume fraction in the three tubes follows,
\begin{eqnarray}
\Phi_A &=& \Phi_\mathrm{br}(\frac{Q_A}{Q_{\mathrm{in},1}},\mathrm{Re}_{\mathrm{in},1} ,\Phi_{{\mathrm{in},1}},\tilde{\mu}_{\mathrm{in},1}) \\
\Phi_C &=& \Phi_\mathrm{run}(\frac{Q_A}{Q_{\mathrm{in},1}},\mathrm{Re}_{\mathrm{in},1}, \Phi_{\mathrm{in},1},\tilde{\mu}_{\mathrm{in},1}) \\
\Phi_B &=& \frac{Q_{\mathrm{in},2} \Phi_{\mathrm{in},2} + Q_C \Phi_C}{Q_B}.
\end{eqnarray}
When the flow in C is negative,
\begin{eqnarray}
\Phi_B &=& \Phi_\mathrm{br}(\frac{Q_B}{Q_{\mathrm{in},2}},\mathrm{Re}_{\mathrm{in},2}, \Phi_{\mathrm{in},2},\tilde{\mu}_{\mathrm{in},2}),\\
\Phi_C &=& \Phi_\mathrm{run}(\frac{Q_B}{Q_{\mathrm{in},2}},\mathrm{Re}_{\mathrm{in},2}, \Phi_{\mathrm{in},2},\tilde{\mu}_{\mathrm{in},2}),\\
\Phi_A &=& \frac{Q_{\mathrm{in},1} \Phi_{\mathrm{in},1} - Q_C \Phi_C}{Q_A}.
\end{eqnarray}
The functions $\Phi_\mathrm{br}$ and $\Phi_\mathrm{run}$ are the phase separation functions which come from either experiment, 3D simulation, or our empirical one-parameter model.
 These functions, as described in Section II, depend upon the relative flow of the branch to the inlet,  the Reynolds number, the inlet volume fraction, and the viscosity contrast between the two fluids.
Once the volume fraction in the tubes is known, the effective viscosity of the network tubes are known from Figure \ref{fig:visc}.
Given that the phase separation functions and resistance are not simple analytical results, a simple result cannot be derived.
The  problem is sufficiently defined to numerically
find the volume fraction and flows through the entire network for fixed inlet conditions.
The above equations provide the relationship for $Q_C/Q_{\mathrm{tot}}$ as a function of $Q_{\mathrm{in},1}/Q_{\mathrm{tot}}$.

\begin{figure}
\begin{center}
a) \includegraphics[width=3.in]{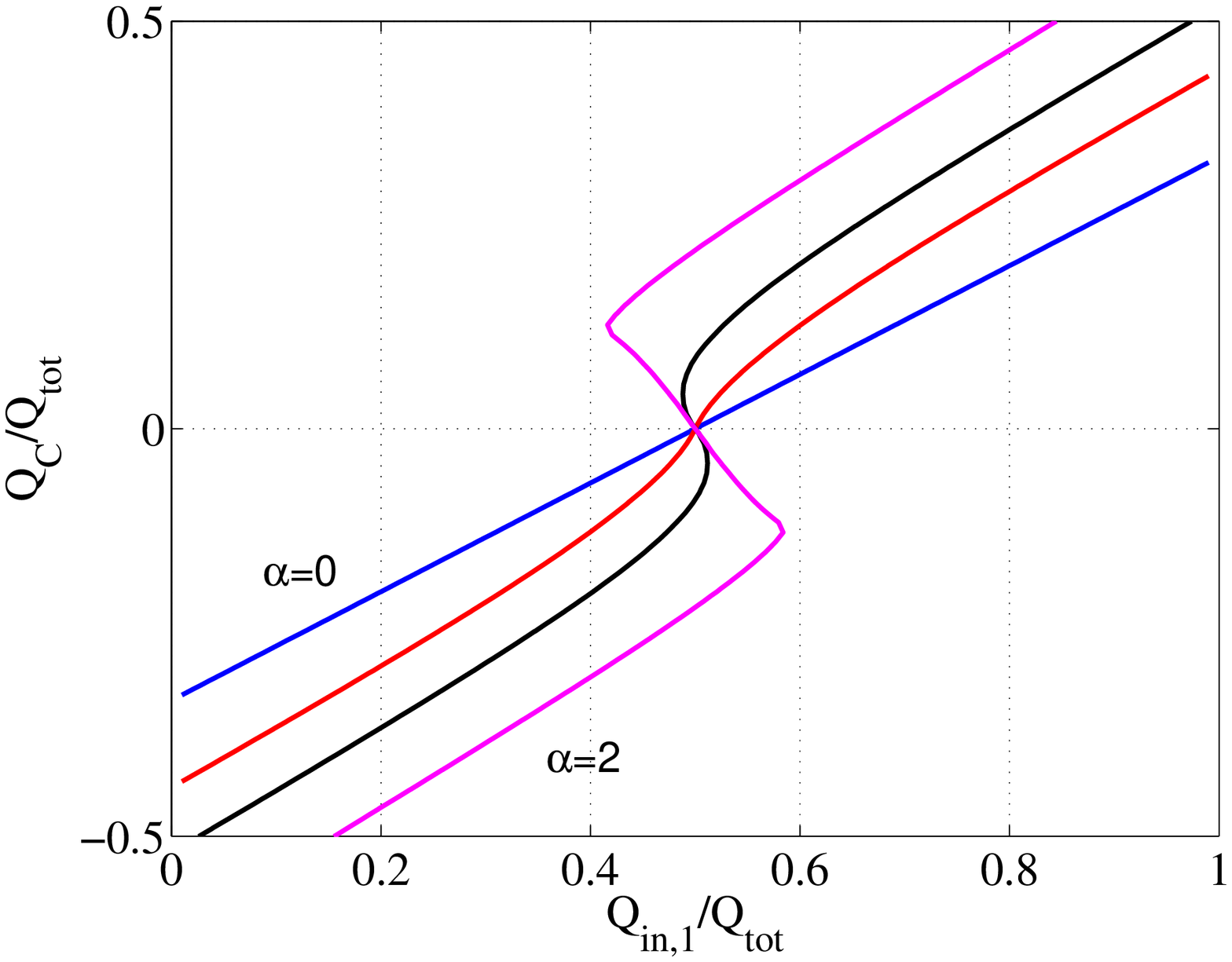}
b) \includegraphics[width=3.in]{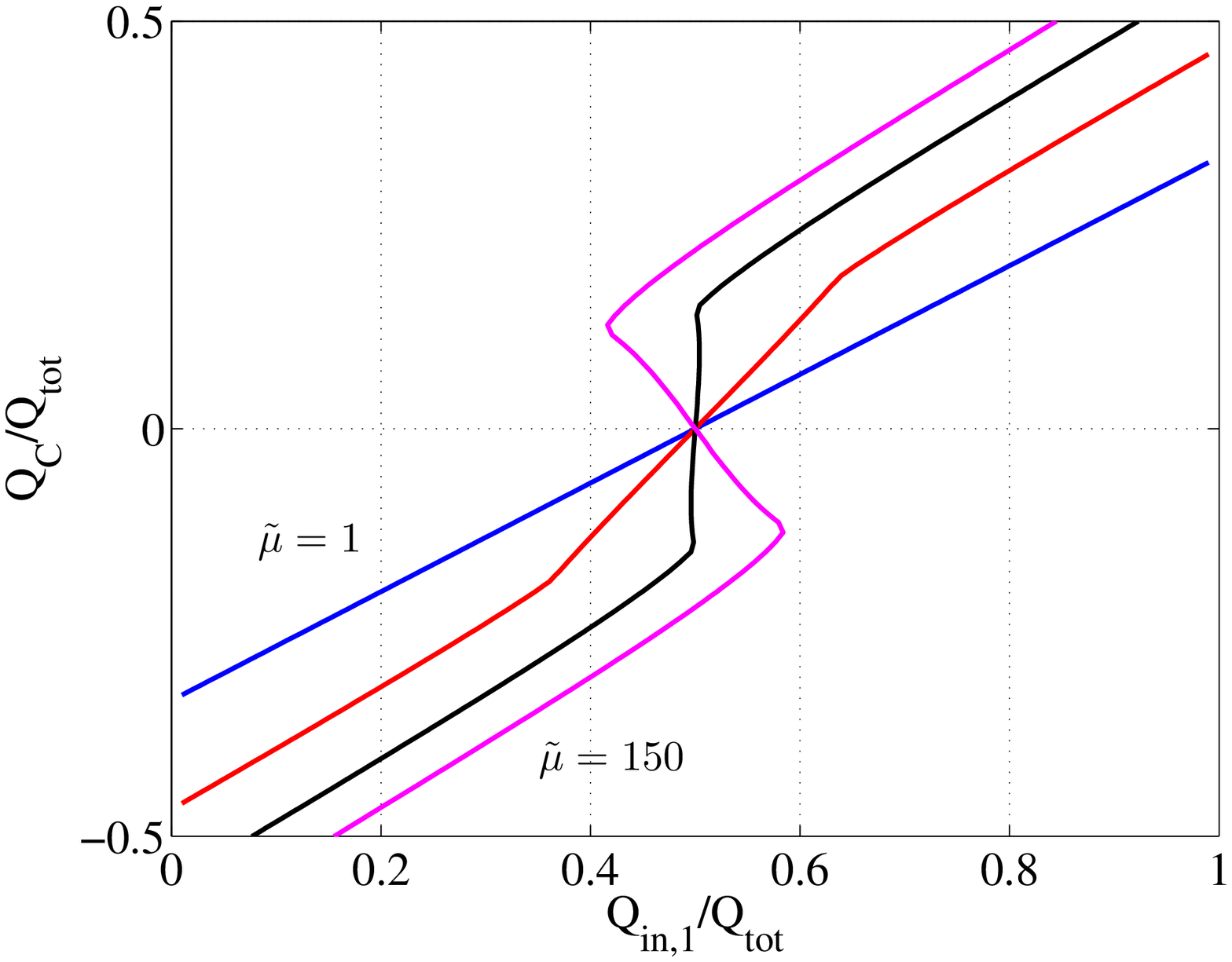}
\end{center}
\caption{a) Predictions of the flow in tube C as a function of the ratio of flow in inlet 1 to the total flow rate $Q_{\mathrm{in},1}/(Q_{\mathrm{in},1}+Q_{\mathrm{in},2})$ for values of $\alpha=0$, $0.5$, $1$, and $2$; (blue, red, black, and magenta respectively). The viscosity contrast is fixed at $\tilde{\mu}=150$ b) Predictions of the flow in tube C for different viscosity contrast of $\tilde{\mu}=1$, $4$, $20$, and $150$; (blue, red, black, and magenta respectively). The parameter for the phase separation function is $\alpha=2$. In both figures a) and b),  the inlet volume fraction is $\Phi_{\mathrm{in},1}=\Phi_{\mathrm{in},2}=0.5$ and the lengths of all tubes in the network are the same.}
\label{fig:Network_predictions}
\end{figure}

Some sample predictions using the one-parameter model are shown in Figure \ref{fig:Network_predictions} for an inlet volume fraction of $\Phi_{\mathrm{in},1}=\Phi_{\mathrm{in},2}=\Phi_\mathrm{in}=0.5$ and a network
where all tubes are equal lengths. In Figure \ref{fig:Network_predictions}a we show the effect of
variation in the one parameter in the phase separation function, $\alpha$, at a high viscosity contrast between the two fluids, $\tilde{\mu}=150$.
It is clear that if sufficiently strong phase separation occurs then bistability in this simple network can emerge. For the curves $\alpha=1$ and $\alpha=2$ there are three possible values of $Q_C$ when $Q_{\mathrm{in},1}/Q_\mathrm{tot}=0.5$. The value at $Q_C=0$ is not stable and thus would not be observed in experiments. While bistability exists for $\alpha=1$, it is much more extreme for larger values of $\alpha$.
In Figure \ref{fig:Network_predictions}b, we show the effect of the  viscosity contrast between the two fluids in the system for a fixed value of $\alpha=2$. We see that the window of bistability grows with the viscosity contrast. While these figures do not span all possible parameters, it is clear that for bistability to emerge in these networks the phase separation  and the viscosity contrast must both be sufficiently strong.

We can easily derive a criterion for the emergence of bistability in this  network flow. We can take the formulation for the network and analyze the derivative of the curve $\partial Q_C/\partial Q_{\mathrm{in},1}$ around the equilibrium point when  $Q_C=0$. At this trivial equilibrium point $Q_{\mathrm{in},1}=Q_{\mathrm{in},2}$ and $Q_A=Q_B$.
Looking at Figure \ref{fig:network_results}, we see bistability emerge when the slope at this point   becomes vertical.
Considering the special case where the two inlet pumps to the network have identical fluid mixtures,  identical  inlet volume fractions, and
 identical tube diameters (i.e. $\Phi_{\mathrm{in},1}=\Phi_{\mathrm{in},2}=\Phi_{\mathrm{in}}$ and $\mu_{\mathrm{in},1}=\mu_{\mathrm{in},2}=\mu_{\mathrm{in}}$),
we obtain the slope of the flow curve to be,
\begin{equation}
\frac{\partial Q_C}{\partial Q_{\mathrm{in},1}}  = \frac{1}{   1 + \frac{\mu_C(Q_C=0)}{\mu_{\mathrm{in}}} \frac{L_C}{L_B + L_A}  -  \frac{1}{\mu_{\mathrm{in}}} \left. \frac{\partial \mu}{ \partial \Phi} \right|_{\Phi_{\mathrm{in}}}( \Phi_{\mathrm{in}} - \Phi_C(Q_C=0))}.
\label{eq:criterion}
\end{equation}
Here, $\mu_C$ and $\Phi_C$ are the viscosity and volume fraction in C when $Q_C=0$.
The value of  $\Phi_C(Q_C=0)$ comes from the phase separation function. For the network configuration shown in Figure \ref{fig:network}, tube C is the run of the T-junction.

We can easily see that bistability would emerge when the denominator of  Equation \ref{eq:criterion} equals zero.
Without loss of generality, we assumed through our definitions that viscosity must increase with volume fraction and thus,
 $\frac{\partial \mu}{ \partial \Phi}>0$.
Since the tube lengths and viscosity must all be positive numbers, Equation \ref{eq:criterion} tells us some general things about when bistability cannot exist.
In the limit of no phase separation, $\Phi_C=\Phi_{\mathrm{in}} $, the denominator would always be positive and bistability could not  exist.
Likewise, if the phase separation is such that $\Phi_C(Q_C=0)>\Phi_{\mathrm{in}}$, then bistability is also not possible.
If  $\Phi_C<\Phi_{{\mathrm{in}}}$ then bistability is possible depending on the network geometry and viscosity of the fluids.
The criterion for the onset of bistability can be stated as,
\begin{equation}
\frac{1}{\mu_{\mathrm{in}}} \left. \frac{\partial \mu}{ \partial \Phi} \right|_{\Phi_{\mathrm{in}}} =
\left(\frac{1 + \frac{\mu_C(Q_C=0)}{\mu_{\mathrm{in}}} \frac{L_C}{L_B + L_A}}{\Phi_{{\mathrm{in}}} - \Phi_C(Q_C=0)} \right).
\label{eq:criterion1}
\end{equation}
To obtain better intuition about when this criterion is met, let us suppose that the
effective viscosity follows a simple law,  $\mu= \tilde{\mu}^\Phi$, where $\tilde{\mu}$ is the viscosity contrast between the two fluids in the system.
With this viscosity law, $\frac{1}{\mu_{\mathrm{in}}} \left. \frac{\partial \mu}{ \partial \Phi} \right|_{\Phi_{\mathrm{in}}} =\mathrm{log}(\tilde{\mu})$.
Therefore the required viscosity contrast between the two fluids, $\tilde{\mu}$, would be exponential in the right hand side of Equation \ref{eq:criterion1}.
While our viscosity law is different it has a similar functional form, thus a small change in the right hand side will get magnified by exponential behavior.

In the limit of $L_C>> (L_A+L_B)$ the right hand side of Equation \ref{eq:criterion1} would grow and thus bistability would become very unlikely.
In the limit when $L_C << (L_A + L_B)$ the criterion would simplify  to,
\[
\frac{1}{\mu_{\mathrm{in}}} \left. \frac{\partial \mu}{ \partial \Phi} \right|_{\Phi_{\mathrm{in}}} = \frac{1} {\Phi_{\mathrm{in}} - \Phi_C(Q_C=0)}.
\]
This value would represent the minimum viscosity contrast needed to observe bistability in any arbitrary network.

We can test Equation \ref{eq:criterion1} against Figure \ref{fig:Network_predictions}b where $\Phi_{\mathrm{in}}=0.5$.
In this figure $L_A=L_B=L_C$ and the phase separation is strong such that $\Phi_C(Q_C=0)=0$.
For these conditions  and our stratified viscosity law, solving our criterion numerically (there is no analytical expression)
would provide a critical viscosity contrast of $\tilde{\mu}=26$. In the Figure \ref{fig:Network_predictions}b
we show a curve for $\tilde{\mu}=20$, and the
slope is nearly vertical, though if we zoom in we find that it is still in the positive direction.

\subsection{Experiment}
A schematic of the experimental network setup is shown in Figure \ref{fig:network}.
We use four inlet pumps to create a controlled stratified flow as the two inlets to the network.
In order to easily monitor the flow rate in the exit branches of the network,
we simply leave exits A and B open to atmospheric  pressure. With this simple setup we  collect the fluid from both outlets  over a fixed time period in order to measure the exit flow rates, $Q_A$ and $Q_B$.
The exit flows are collected in a disposable test tube and the fluid mass  is weighed on an analytical balance.

In the network experiments, we used the same T-junction geometry described in Section II. However, in these experiments after the T-junction
the tubes A, B, and C were narrowed down to tubing of 0.5 mm inner diameter. This decrease in diameter
was done to increase the hydraulic resistance between the network connections. Since the outlets A and B were held open at atmospheric pressure, we want the hydraulic resistance of the tubes to significantly dominate over surface tension effects due to dripping at the exits or hydrostatic pressure differences due to A and B not being at precisely the same height. In all experiments reported here the  hydraulic resistance of A, B, and C were the same.

 In these experiments the pumps on the left and right side are always kept in the same proportion such that the volume fraction in the two inlets are constant at $\Phi_{\mathrm{in},1}=\Phi_{\mathrm{in},2}$.
In the experiment the pumps on inlet 1 in Figure \ref{fig:network} are held at a constant flow rate of 2 ml/min.
The total flow rate in inlet 2 is varied between 8 and 0.5 ml/min in order to change the inlet network flow ratio, $Q_{\mathrm{in},1}/Q_{\mathrm{tot}}$.
Each experimental data point is repeated in three independent trials where new networks are constructed and  new solutions mixed.

We used our model phase separation function to help guide us to regimes where strong bistability would be expected.
We determined from our experiments that strong phase separation occurs when
the inlet volume fraction is  30\%, see Figure \ref{fig:fitfcn}a. We also  determined from our  network analysis
that a viscosity contrast of 150 would give us a strong window of bistability in the network;  Figure \ref{fig:Network_predictions}b.
Since we found in earlier sections that the amount of phase separation is not very sensitive to the viscosity contrast, we assume that the
fit value of $\alpha=2.2$ that was found in Figure \ref{fig:fitfcn}a for a viscosity contrast of 15 is approximately equivalent to that at a contrast of 150.
 We  use the fit  function with $\alpha=2.2$ in our network prediction.

\subsection{Results}

\begin{figure}
\begin{center}
a) \includegraphics[width=3.in]{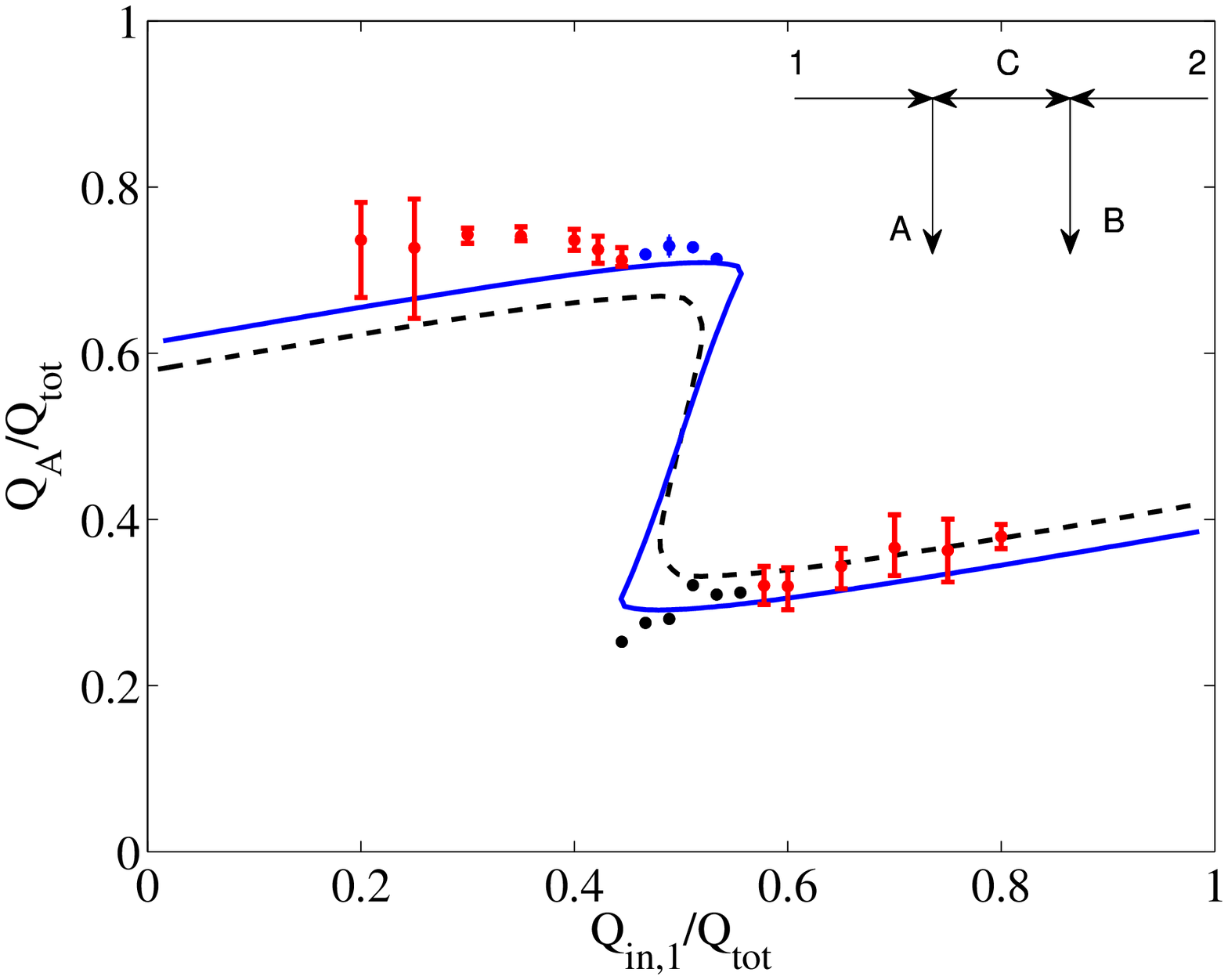}
b) \includegraphics[width=3.in]{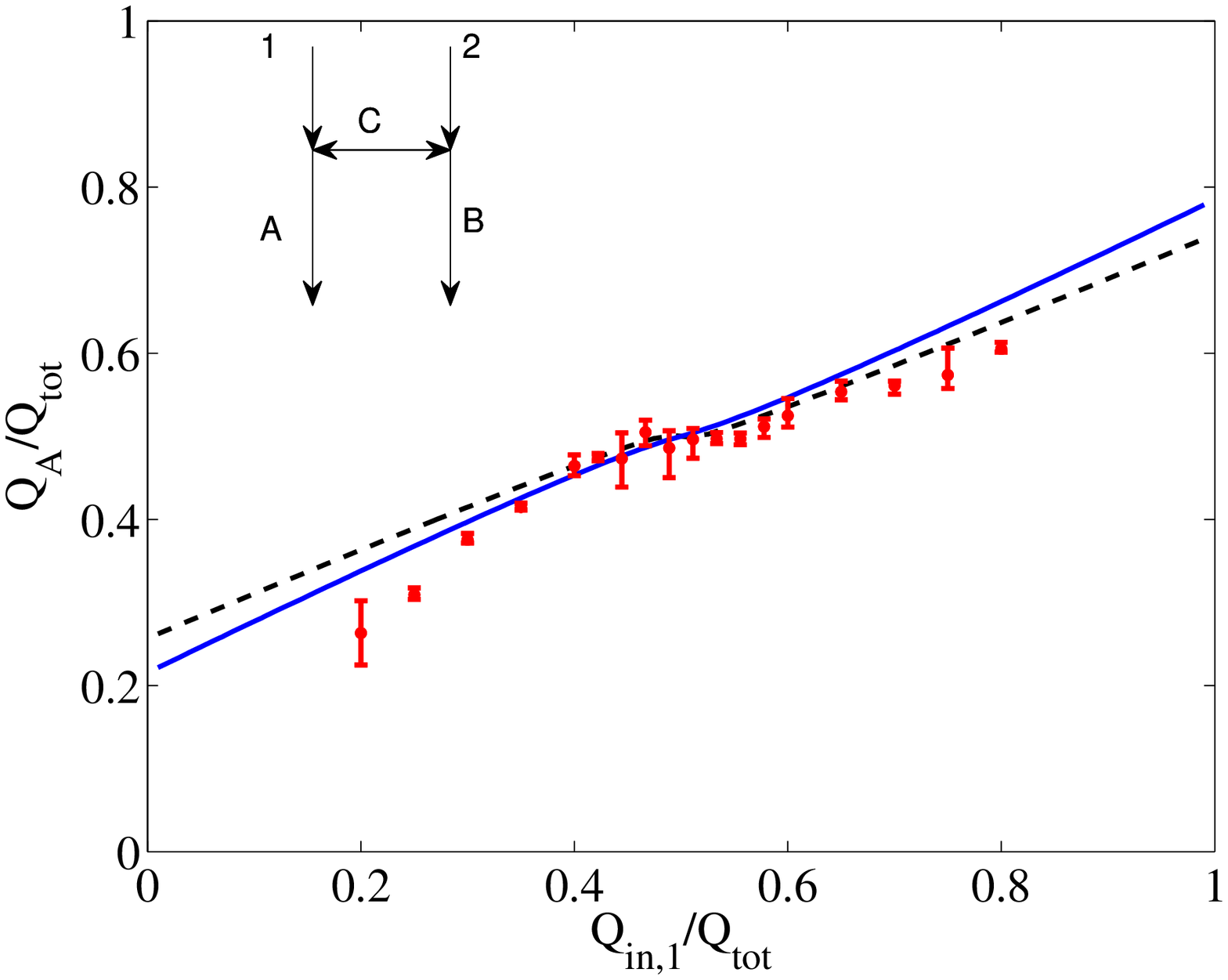}
\end{center}
\caption{a) Flow in exit branch A of the network in Figure \ref{fig:network}b determined from experiment and theory. Notice the region of bistability around the
state where the flow rates on the inlets are equal. Both inlets are a 30\% sucrose and 70\% water mixture in stratified flow. The viscosity contrast between the sucrose solution in inlet pumps and water  is  150. The inset shows the network topology.
The error bars on the red experimental data points show the maximum and minimum value recorded in three independent trials and the data points shows the mean.
The blue data points (and error bars) were taken on two of the trials where we proceeded from left to right along the flow curve. The black data points were from one trial where we moved from right to left.
In b) we have the exact same setup, only the connection for the network is made by the branches of the two inlet T-junctions - see the inset.
In both figures the points are experimental data, the solid line is from the theory, using the fit function with $\alpha=2.2$ and the dashed line is from the theory using the  phase separation as predicted from the Comsol simulation.
  }
\label{fig:network_results}
\end{figure}

The predictions of the network model compared to our experiments are shown in Figure \ref{fig:network_results}a.
We show the prediction using the simple phase separation fit function; shown as the solid line.
To further validate our predictions,  we also used our finite element simulation at the experimental parameter values and
 get the predicted phase separation function as well (the result is very similar to Figure \ref{fig:phi_in}b).
The network prediction based on the
phase separation function calculated using the 3D finite element code is shown as the  dashed line.
A clear window of bistability exists and the behavior is quite remarkable. If we slowly adjust the incoming flow rates around the point where $Q_{\mathrm{in},1}=Q_{\mathrm{in},2}$ then the flow in the outlet undergoes a dramatic jump when we pass the transition point. The agreement between experiment and theory is excellent. Note that the window of bistability predicted by the model using the phase separation which came from the finite element simulation is not as strong as we find experimentally. This result is consistent with what we found in  an earlier section where the finite element simulation typically under-predicts the amount of phase separation seen in experiment.

In the experiment the pumps on inlet 1 in Figure \ref{fig:network} are held at a constant flow rate of 2 ml/min.
When the system is in a state such that the flow in C is positive (generally $Q_{\mathrm{in},1}/Q_{\mathrm{tot}}>0.5$) then phase separation only occurs at the T-junction on inlet 1. Since this flow rate is constant, we expect a single value of $\alpha$ to suffice for describing the behavior.
When the flow is in a state where the flow in branch C is negative (generally $Q_{\mathrm{in},1}/Q_{\mathrm{tot}}<0.5$) then phase separation only occurs at the T-junction between branches B and C. In order to span values of $Q_{\mathrm{in},1}/Q_{\mathrm{tot}}$,
the pumps on Inlet 2 are varied from an inlet flow rate of 8 to 0.5 ml/min in the experiment.
Since the phase separation function is sensitive to the total flow rate, then the value of $\alpha$ would be expected to be variable. The experimental data show an asymmetry about   $Q_{\mathrm{in},1}/Q_{\mathrm{tot}}=0.5$ which is not present in the simple model which assumes constant phase separation behavior. Despite these differences, the agreement between the data and the prediction is excellent.
Note the  phase separation function calculated with the 3D finite element simulation was run at a constant total flow rate of 4 ml/min to approximately correspond to the average total flow in the network across the range of parameters.

Another interesting feature of this system is that if we keep the same basic network but reconfigure the system such that tube C in the network is the branch of both inlet T-junctions rather  than the run, the window of bistability disappears completely.
These data and the predictions are shown in Figure \ref{fig:network_results}b.
The experimental data are consistent with the model. Again the data show an asymmetry about $Q_{\mathrm{in,1}}/Q_{\mathrm{tot}}$ which is not present in the model due to our assumption of constant phase separation for all inlet flow rates.

An interesting feature of Figure \ref{fig:network_results}b, is that both the data and the prediction using the phase separation from the finite element simulation show a flattening of the slope around the equilibrium point, $Q_C=0$, that is not seen with the simple fit function. The reason is that the fit function assumes that the volume fraction in the branch goes to $\Phi_{\mathrm{in}}$ when the flow in the branch is zero; ie. $\Phi_{\mathrm{br}}(Q_{\mathrm{br}}=0)=\Phi_{\mathrm{in}}=0.3$. This assumption is made to keep the fit function to one adjustable parameter, the assumption  is not based on data or some physical mechanism. As explained in Section II, the experimental setup does not allow us to accurately determine the branch volume fraction at this point.  However, the phase separation functions computed by the finite element simulations show that the volume fraction in the branch as the flow in the branch goes to zero is likely somewhat less than the inlet volume fraction. For the case here, the finite element simulation predicts $\Phi_{\mathrm{br}}(Q_{\mathrm{br}}=0)\approx 0.18$. Since the flow in tube C is close to zero when $Q_{\mathrm{in},1}/Q_{\mathrm{tot}}\approx0.5$, it is the phase separation behavior when the branch flow is close to zero which is important.

What is interesting about the dashed curves in Figure \ref{fig:network_results} is that the prediction is based entirely on first principles and simulation. The complex phase separation functions and effective viscosity laws  are calculated with a Navier Stokes simulation. The dashed curves in these figures have no fit parameters or assumptions inherent in them. The solid curve is based on a simple one parameter phase separation model that is convenient for exploring behavior and parameter space, but it should not be taken as more than a fit function.

Finally, the criterion in Equation \ref{eq:criterion1} provides a simple explanation as to why bistability is seen in one network configuration but not the other.
In the configuration of the network shown in Figure \ref{fig:network_results}a,
tube C is the run of the inlet T-junctions and thus $\Phi_C(Q_C=0) \approx 0 $.
Under these conditions bistability is possible when $\tilde{\mu}>50$; our viscosity contrast of 150 exceeds the minimum  value.
For the configuration where the network is connected such that tube C is the branch of the T-junction,
$\Phi_C(Q_C=0) = \Phi_{\mathrm{in}} $ if we use the fit function; bistability is not possible under this condition. If we use the value of $\Phi_C(Q_C=0) = 0.18$ which is computed in the finite element simulation, then the required viscosity contrast would be   $\tilde{\mu}>5900$.

\section{Conclusions}
In this work,  we have demonstrated that  laminar flow of two miscible fluids of different viscosity approaching a T-junction has  significant phase separation.
We measured phase distribution functions for this system for the first time (to the best of our knowledge). We compared our experiments on these phase distribution functions to 3D Navier Stokes simulations and find excellent agreement. We find that in the geometry used in this work, that inertia was the mechanism behind the phase separation. We described the phase separation as a function of the key parameters such as Reynolds number and viscosity contrast between the two fluids.

Once such phase separation functions were known, we  explored  the consequences  for phase distribution within a simple network.
We find theoretically and experimentally that the unequal separation of two phases at a single
bifurcation can lead to multiple equilibrium states  in a network.
We derive a
 criterion for the existence of multiple  states  and found  this criterion to
 explain  the experimentally observed behavior.
The phase separation functions depend on the exact details of the system;
the fluids used and the geometry of the network bifurcation.
However, the network results are generic and could be applied to or found in different systems.

\section*{Acknowledgements}
This work was supported by the  National Science Foundation under contract DMS-1211640.

\bibliography{refs_network}

\begin{thebibliography}{36}%
\makeatletter
\providecommand \@ifxundefined [1]{%
 \@ifx{#1\undefined}
}%
\providecommand \@ifnum [1]{%
 \ifnum #1\expandafter \@firstoftwo
 \else \expandafter \@secondoftwo
 \fi
}%
\providecommand \@ifx [1]{%
 \ifx #1\expandafter \@firstoftwo
 \else \expandafter \@secondoftwo
 \fi
}%
\providecommand \natexlab [1]{#1}%
\providecommand \enquote  [1]{``#1''}%
\providecommand \bibnamefont  [1]{#1}%
\providecommand \bibfnamefont [1]{#1}%
\providecommand \citenamefont [1]{#1}%
\providecommand \href@noop [0]{\@secondoftwo}%
\providecommand \href [0]{\begingroup \@sanitize@url \@href}%
\providecommand \@href[1]{\@@startlink{#1}\@@href}%
\providecommand \@@href[1]{\endgroup#1\@@endlink}%
\providecommand \@sanitize@url [0]{\catcode `\\12\catcode `\$12\catcode
  `\&12\catcode `\#12\catcode `\^12\catcode `\_12\catcode `\%12\relax}%
\providecommand \@@startlink[1]{}%
\providecommand \@@endlink[0]{}%
\providecommand \url  [0]{\begingroup\@sanitize@url \@url }%
\providecommand \@url [1]{\endgroup\@href {#1}{\urlprefix }}%
\providecommand \urlprefix  [0]{URL }%
\providecommand \Eprint [0]{\href }%
\providecommand \doibase [0]{http://dx.doi.org/}%
\providecommand \selectlanguage [0]{\@gobble}%
\providecommand \bibinfo  [0]{\@secondoftwo}%
\providecommand \bibfield  [0]{\@secondoftwo}%
\providecommand \translation [1]{[#1]}%
\providecommand \BibitemOpen [0]{}%
\providecommand \bibitemStop [0]{}%
\providecommand \bibitemNoStop [0]{.\EOS\space}%
\providecommand \EOS [0]{\spacefactor3000\relax}%
\providecommand \BibitemShut  [1]{\csname bibitem#1\endcsname}%
\let\auto@bib@innerbib\@empty
\bibitem [{\citenamefont {Jousse}\ \emph {et~al.}(2006)\citenamefont {Jousse},
  \citenamefont {Farr}, \citenamefont {Link}, \citenamefont {Fuerstman},\ and\
  \citenamefont {Garstecki}}]{Jousse2006}%
  \BibitemOpen
  \bibfield  {author} {\bibinfo {author} {\bibfnamefont {F.}~\bibnamefont
  {Jousse}}, \bibinfo {author} {\bibfnamefont {R.}~\bibnamefont {Farr}},
  \bibinfo {author} {\bibfnamefont {D.~R.}\ \bibnamefont {Link}}, \bibinfo
  {author} {\bibfnamefont {M.~J.}\ \bibnamefont {Fuerstman}}, \ and\ \bibinfo
  {author} {\bibfnamefont {P.}~\bibnamefont {Garstecki}},\ }\bibfield  {title}
  {\enquote {\bibinfo {title} {Bifurcation of droplet flows within
  capillaries},}\ }\href@noop {} {\bibfield  {journal} {\bibinfo  {journal}
  {Phys. Rev. E}\ }\textbf {\bibinfo {volume} {74}},\ \bibinfo {pages} {036311}
  (\bibinfo {year} {2006})}\BibitemShut {NoStop}%
\bibitem [{\citenamefont {Schindler}\ and\ \citenamefont
  {Ajdari}(2008)}]{Schindler2008}%
  \BibitemOpen
  \bibfield  {author} {\bibinfo {author} {\bibfnamefont {M.}~\bibnamefont
  {Schindler}}\ and\ \bibinfo {author} {\bibfnamefont {A.}~\bibnamefont
  {Ajdari}},\ }\bibfield  {title} {\enquote {\bibinfo {title} {Droplet traffic
  in microfluidic networks: A simple model for understanding and designing},}\
  }\href@noop {} {\bibfield  {journal} {\bibinfo  {journal} {Phys. Rev. Lett.}\
  }\textbf {\bibinfo {volume} {100}},\ \bibinfo {pages} {1--4} (\bibinfo {year}
  {2008})}\BibitemShut {NoStop}%
\bibitem [{\citenamefont {Fuerstman}\ \emph {et~al.}(2007)\citenamefont
  {Fuerstman}, \citenamefont {Garstecki},\ and\ \citenamefont
  {Whitesides}}]{Fuerstman2007}%
  \BibitemOpen
  \bibfield  {author} {\bibinfo {author} {\bibfnamefont {M.~J.}\ \bibnamefont
  {Fuerstman}}, \bibinfo {author} {\bibfnamefont {P.}~\bibnamefont
  {Garstecki}}, \ and\ \bibinfo {author} {\bibfnamefont {G.~M.}\ \bibnamefont
  {Whitesides}},\ }\bibfield  {title} {\enquote {\bibinfo {title}
  {Coding/decoding and reversibility of droplet trains in microfluidic
  networks},}\ }\href@noop {} {\bibfield  {journal} {\bibinfo  {journal}
  {Science}\ }\textbf {\bibinfo {volume} {315}},\ \bibinfo {pages} {828--32}
  (\bibinfo {year} {2007})}\BibitemShut {NoStop}%
\bibitem [{\citenamefont {Prakash}\ and\ \citenamefont
  {Gershenfeld}(2007)}]{Prakash2007}%
  \BibitemOpen
  \bibfield  {author} {\bibinfo {author} {\bibfnamefont {M.}~\bibnamefont
  {Prakash}}\ and\ \bibinfo {author} {\bibfnamefont {N.}~\bibnamefont
  {Gershenfeld}},\ }\bibfield  {title} {\enquote {\bibinfo {title}
  {Microfluidic bubble logic},}\ }\href@noop {} {\bibfield  {journal} {\bibinfo
   {journal} {Science}\ }\textbf {\bibinfo {volume} {315}},\ \bibinfo {pages}
  {832--5} (\bibinfo {year} {2007})}\BibitemShut {NoStop}%
\bibitem [{\citenamefont {Joanicot}\ and\ \citenamefont
  {Ajdari}(2005)}]{Joanicot2005}%
  \BibitemOpen
  \bibfield  {author} {\bibinfo {author} {\bibfnamefont {M.}~\bibnamefont
  {Joanicot}}\ and\ \bibinfo {author} {\bibfnamefont {A.}~\bibnamefont
  {Ajdari}},\ }\bibfield  {title} {\enquote {\bibinfo {title} {Droplet control
  for microfluidics},}\ }\href@noop {} {\bibfield  {journal} {\bibinfo
  {journal} {Science}\ }\textbf {\bibinfo {volume} {309}},\ \bibinfo {pages}
  {887--888} (\bibinfo {year} {2005})}\BibitemShut {NoStop}%
\bibitem [{\citenamefont {Helfrich}(1995)}]{Helfrich1995}%
  \BibitemOpen
  \bibfield  {author} {\bibinfo {author} {\bibfnamefont {K.~R.}\ \bibnamefont
  {Helfrich}},\ }\bibfield  {title} {\enquote {\bibinfo {title} {Thermo-viscous
  fingering of flow in a thin gap: a model of magma flow in dikes and
  fissures},}\ }\href@noop {} {\bibfield  {journal} {\bibinfo  {journal}
  {Journal of Fluid Mechanics}\ }\textbf {\bibinfo {volume} {305}},\ \bibinfo
  {pages} {219--238} (\bibinfo {year} {1995})}\BibitemShut {NoStop}%
\bibitem [{\citenamefont {Wylie}\ \emph {et~al.}(1999)\citenamefont {Wylie},
  \citenamefont {Voight},\ and\ \citenamefont {Whitehead}}]{Wylie1999}%
  \BibitemOpen
  \bibfield  {author} {\bibinfo {author} {\bibfnamefont {J.~J.}\ \bibnamefont
  {Wylie}}, \bibinfo {author} {\bibfnamefont {B.}~\bibnamefont {Voight}}, \
  and\ \bibinfo {author} {\bibfnamefont {J.~A.}\ \bibnamefont {Whitehead}},\
  }\bibfield  {title} {\enquote {\bibinfo {title} {Instability of magma flow
  from volatile-dependent viscosity},}\ }\href@noop {} {\bibfield  {journal}
  {\bibinfo  {journal} {Science}\ }\textbf {\bibinfo {volume} {285}},\ \bibinfo
  {pages} {1883--1885} (\bibinfo {year} {1999})}\BibitemShut {NoStop}%
\bibitem [{\citenamefont {Krogh}(1921)}]{Krogh:1921aa}%
  \BibitemOpen
  \bibfield  {author} {\bibinfo {author} {\bibfnamefont {A.}~\bibnamefont
  {Krogh}},\ }\bibfield  {title} {\enquote {\bibinfo {title} {Studies on the
  physiology of capillaries: {II}. the reactions to local stimuli of the
  blood-vessels in the skin and web of the frog},}\ }\href@noop {} {\bibfield
  {journal} {\bibinfo  {journal} {J Physiol (Lond)}\ }\textbf {\bibinfo
  {volume} {55}},\ \bibinfo {pages} {412--22} (\bibinfo {year}
  {1921})}\BibitemShut {NoStop}%
\bibitem [{\citenamefont {Krogh}(1922)}]{Krogh:1922aa}%
  \BibitemOpen
  \bibfield  {author} {\bibinfo {author} {\bibfnamefont {A.}~\bibnamefont
  {Krogh}},\ }\href@noop {} {\emph {\bibinfo {title} {The Anatomy and
  Physiology of Capillaries}}}\ (\bibinfo  {publisher} {Yale University
  Press},\ \bibinfo {year} {1922})\BibitemShut {NoStop}%
\bibitem [{\citenamefont {Backer}\ \emph {et~al.}(2002)\citenamefont {Backer},
  \citenamefont {Creteur}, \citenamefont {Preiser}, \citenamefont {Dubois},\
  and\ \citenamefont {Vincent}}]{debacker:2002}%
  \BibitemOpen
  \bibfield  {author} {\bibinfo {author} {\bibfnamefont {D.~De}\ \bibnamefont
  {Backer}}, \bibinfo {author} {\bibfnamefont {J.}~\bibnamefont {Creteur}},
  \bibinfo {author} {\bibfnamefont {J.~C.}\ \bibnamefont {Preiser}}, \bibinfo
  {author} {\bibfnamefont {M.~J.}\ \bibnamefont {Dubois}}, \ and\ \bibinfo
  {author} {\bibfnamefont {J.~L.}\ \bibnamefont {Vincent}},\ }\bibfield
  {title} {\enquote {\bibinfo {title} {Microvascular blood flow is altered in
  patients with sepsis},}\ }\href@noop {} {\bibfield  {journal} {\bibinfo
  {journal} {Am. J. of Respitory and Critical Care Med.}\ }\textbf {\bibinfo
  {volume} {166}},\ \bibinfo {pages} {98--104} (\bibinfo {year}
  {2002})}\BibitemShut {NoStop}%
\bibitem [{\citenamefont {Rodgers}\ \emph {et~al.}(1984)\citenamefont
  {Rodgers}, \citenamefont {Schechter}, \citenamefont {Noguchi}, \citenamefont
  {Klein}, \citenamefont {Niehuis},\ and\ \citenamefont {Bonner}}]{rodgers}%
  \BibitemOpen
  \bibfield  {author} {\bibinfo {author} {\bibfnamefont {G.~P.}\ \bibnamefont
  {Rodgers}}, \bibinfo {author} {\bibfnamefont {A.~N.}\ \bibnamefont
  {Schechter}}, \bibinfo {author} {\bibfnamefont {C.~T.}\ \bibnamefont
  {Noguchi}}, \bibinfo {author} {\bibfnamefont {H.~G.}\ \bibnamefont {Klein}},
  \bibinfo {author} {\bibfnamefont {Q.~W.}\ \bibnamefont {Niehuis}}, \ and\
  \bibinfo {author} {\bibfnamefont {R.~F.}\ \bibnamefont {Bonner}},\ }\bibfield
   {title} {\enquote {\bibinfo {title} {Periodic microcirculatory flow in
  patients with sickle cell disease},}\ }\href@noop {} {\bibfield  {journal}
  {\bibinfo  {journal} {New England J. of Medicine}\ }\textbf {\bibinfo
  {volume} {311}},\ \bibinfo {pages} {1534--1538} (\bibinfo {year}
  {1984})}\BibitemShut {NoStop}%
\bibitem [{\citenamefont {Kiani}\ \emph {et~al.}(1994)\citenamefont {Kiani},
  \citenamefont {Pries}, \citenamefont {Hsu}, \citenamefont {Sarelius},\ and\
  \citenamefont {Cokelet}}]{Kiani:1994aa}%
  \BibitemOpen
  \bibfield  {author} {\bibinfo {author} {\bibfnamefont {M.~F.}\ \bibnamefont
  {Kiani}}, \bibinfo {author} {\bibfnamefont {A.~R.}\ \bibnamefont {Pries}},
  \bibinfo {author} {\bibfnamefont {L.~L.}\ \bibnamefont {Hsu}}, \bibinfo
  {author} {\bibfnamefont {I.~H.}\ \bibnamefont {Sarelius}}, \ and\ \bibinfo
  {author} {\bibfnamefont {G.~R.}\ \bibnamefont {Cokelet}},\ }\bibfield
  {title} {\enquote {\bibinfo {title} {Fluctuations in microvascular blood flow
  parameters caused by hemodynamic mechanisms},}\ }\href@noop {} {\bibfield
  {journal} {\bibinfo  {journal} {Am J Physiol}\ }\textbf {\bibinfo {volume}
  {266}},\ \bibinfo {pages} {H1822--8} (\bibinfo {year} {1994})}\BibitemShut
  {NoStop}%
\bibitem [{\citenamefont {Carr}\ and\ \citenamefont
  {Lacoin}(2000)}]{Carr:2000aa}%
  \BibitemOpen
  \bibfield  {author} {\bibinfo {author} {\bibfnamefont {R.~T.}\ \bibnamefont
  {Carr}}\ and\ \bibinfo {author} {\bibfnamefont {M.}~\bibnamefont {Lacoin}},\
  }\bibfield  {title} {\enquote {\bibinfo {title} {Nonlinear dynamics of
  microvascular blood flow},}\ }\href@noop {} {\bibfield  {journal} {\bibinfo
  {journal} {Annals of biomedical engineering}\ }\textbf {\bibinfo {volume}
  {28}},\ \bibinfo {pages} {641--52} (\bibinfo {year} {2000})}\BibitemShut
  {NoStop}%
\bibitem [{\citenamefont {Geddes}\ \emph {et~al.}(2007)\citenamefont {Geddes},
  \citenamefont {Carr}, \citenamefont {Karst},\ and\ \citenamefont
  {Wu}}]{Geddes:2007}%
  \BibitemOpen
  \bibfield  {author} {\bibinfo {author} {\bibfnamefont {J.~B.}\ \bibnamefont
  {Geddes}}, \bibinfo {author} {\bibfnamefont {R.~T.}\ \bibnamefont {Carr}},
  \bibinfo {author} {\bibfnamefont {N.}~\bibnamefont {Karst}}, \ and\ \bibinfo
  {author} {\bibfnamefont {F.}~\bibnamefont {Wu}},\ }\bibfield  {title}
  {\enquote {\bibinfo {title} {The onset of oscillations in microvascular blood
  flow},}\ }\href@noop {} {\bibfield  {journal} {\bibinfo  {journal} {SIAM
  Journal on Applied Dynamical Systems}\ }\textbf {\bibinfo {volume} {6}},\
  \bibinfo {pages} {694--727} (\bibinfo {year} {2007})}\BibitemShut {NoStop}%
\bibitem [{\citenamefont {hr\ae us}\ and\ \citenamefont
  {Lindqvist}(1931)}]{Fahraeus:1931aa}%
  \BibitemOpen
  \bibfield  {author} {\bibinfo {author} {\bibfnamefont {R.~F\aa}\ \bibnamefont
  {hr\ae us}}\ and\ \bibinfo {author} {\bibfnamefont {T.}~\bibnamefont
  {Lindqvist}},\ }\bibfield  {title} {\enquote {\bibinfo {title} {The viscosity
  of blood in narrow capillary tubes},}\ }\href@noop {} {\bibfield  {journal}
  {\bibinfo  {journal} {Journal of Physiology}\ }\textbf {\bibinfo {volume}
  {96}},\ \bibinfo {pages} {562--568} (\bibinfo {year} {1931})}\BibitemShut
  {NoStop}%
\bibitem [{\citenamefont {Geddes}\ \emph {et~al.}(2010)\citenamefont {Geddes},
  \citenamefont {Storey}, \citenamefont {Gardner},\ and\ \citenamefont
  {Carr}}]{geddes2010}%
  \BibitemOpen
  \bibfield  {author} {\bibinfo {author} {\bibfnamefont {J.~G.}\ \bibnamefont
  {Geddes}}, \bibinfo {author} {\bibfnamefont {B.~D.}\ \bibnamefont {Storey}},
  \bibinfo {author} {\bibfnamefont {D.}~\bibnamefont {Gardner}}, \ and\
  \bibinfo {author} {\bibfnamefont {R.~T.}\ \bibnamefont {Carr}},\ }\bibfield
  {title} {\enquote {\bibinfo {title} {Bistability in a simple fluid network
  due to viscosity contrast},}\ }\href@noop {} {\bibfield  {journal} {\bibinfo
  {journal} {Phys. Rev. E}\ }\textbf {\bibinfo {volume} {81}},\ \bibinfo
  {pages} {046316} (\bibinfo {year} {2010})}\BibitemShut {NoStop}%
\bibitem [{\citenamefont {Bugliarello}\ and\ \citenamefont
  {Hsiao}(1964)}]{Bugliarello:1964aa}%
  \BibitemOpen
  \bibfield  {author} {\bibinfo {author} {\bibfnamefont {G.}~\bibnamefont
  {Bugliarello}}\ and\ \bibinfo {author} {\bibfnamefont {C.~C.}\ \bibnamefont
  {Hsiao}},\ }\bibfield  {title} {\enquote {\bibinfo {title} {Phase separation
  in suspensions flowing through bifurcations: A simplified hemodynamic
  model},}\ }\href@noop {} {\bibfield  {journal} {\bibinfo  {journal}
  {Science}\ }\textbf {\bibinfo {volume} {143}},\ \bibinfo {pages} {469--71}
  (\bibinfo {year} {1964})}\BibitemShut {NoStop}%
\bibitem [{\citenamefont {Chien}\ \emph {et~al.}(1985)\citenamefont {Chien},
  \citenamefont {Tvetenstrand}, \citenamefont {Epstein},\ and\ \citenamefont
  {Schmid-Sch{\"o}nbein}}]{Chien:1985aa}%
  \BibitemOpen
  \bibfield  {author} {\bibinfo {author} {\bibfnamefont {S.}~\bibnamefont
  {Chien}}, \bibinfo {author} {\bibfnamefont {C.~D.}\ \bibnamefont
  {Tvetenstrand}}, \bibinfo {author} {\bibfnamefont {M.~A.}\ \bibnamefont
  {Epstein}}, \ and\ \bibinfo {author} {\bibfnamefont {G.~W.}\ \bibnamefont
  {Schmid-Sch{\"o}nbein}},\ }\bibfield  {title} {\enquote {\bibinfo {title}
  {Model studies on distributions of blood cells at microvascular
  bifurcations},}\ }\href@noop {} {\bibfield  {journal} {\bibinfo  {journal}
  {Am J Physiol}\ }\textbf {\bibinfo {volume} {248}},\ \bibinfo {pages}
  {H568--76} (\bibinfo {year} {1985})}\BibitemShut {NoStop}%
\bibitem [{\citenamefont {Dellimore}\ \emph {et~al.}(1983)\citenamefont
  {Dellimore}, \citenamefont {Dunlop},\ and\ \citenamefont
  {Canham}}]{Dellimore:1983aa}%
  \BibitemOpen
  \bibfield  {author} {\bibinfo {author} {\bibfnamefont {J.~W.}\ \bibnamefont
  {Dellimore}}, \bibinfo {author} {\bibfnamefont {M.~J.}\ \bibnamefont
  {Dunlop}}, \ and\ \bibinfo {author} {\bibfnamefont {P.~B.}\ \bibnamefont
  {Canham}},\ }\bibfield  {title} {\enquote {\bibinfo {title} {Ratio of cells
  and plasma in blood flowing past branches in small plastic channels},}\
  }\href@noop {} {\bibfield  {journal} {\bibinfo  {journal} {Am J Physiol}\
  }\textbf {\bibinfo {volume} {244}},\ \bibinfo {pages} {H635--43} (\bibinfo
  {year} {1983})}\BibitemShut {NoStop}%
\bibitem [{\citenamefont {Fenton}\ \emph {et~al.}(1985)\citenamefont {Fenton},
  \citenamefont {Carr},\ and\ \citenamefont {Cokelet}}]{Fenton:1985aa}%
  \BibitemOpen
  \bibfield  {author} {\bibinfo {author} {\bibfnamefont {B.~M.}\ \bibnamefont
  {Fenton}}, \bibinfo {author} {\bibfnamefont {R.~T.}\ \bibnamefont {Carr}}, \
  and\ \bibinfo {author} {\bibfnamefont {G.~R.}\ \bibnamefont {Cokelet}},\
  }\bibfield  {title} {\enquote {\bibinfo {title} {Nonuniform red cell
  distribution in 20 to 100 micrometers bifurcations},}\ }\href@noop {}
  {\bibfield  {journal} {\bibinfo  {journal} {Microvasc Res}\ }\textbf
  {\bibinfo {volume} {29}},\ \bibinfo {pages} {103--26} (\bibinfo {year}
  {1985})}\BibitemShut {NoStop}%
\bibitem [{\citenamefont {Klitzman}\ and\ \citenamefont
  {Johnson}(1982)}]{Klitzman:1982aa}%
  \BibitemOpen
  \bibfield  {author} {\bibinfo {author} {\bibfnamefont {B.}~\bibnamefont
  {Klitzman}}\ and\ \bibinfo {author} {\bibfnamefont {P.~C.}\ \bibnamefont
  {Johnson}},\ }\bibfield  {title} {\enquote {\bibinfo {title} {Capillary
  network geometry and red cell distribution in hamster cremaster muscle},}\
  }\href@noop {} {\bibfield  {journal} {\bibinfo  {journal} {Am J Physiol}\
  }\textbf {\bibinfo {volume} {242}},\ \bibinfo {pages} {H211--9} (\bibinfo
  {year} {1982})}\BibitemShut {NoStop}%
\bibitem [{\citenamefont {Pries}\ \emph {et~al.}(1989)\citenamefont {Pries},
  \citenamefont {Ley}, \citenamefont {Claassen},\ and\ \citenamefont
  {Gaehtgens}}]{Pries:1989aa}%
  \BibitemOpen
  \bibfield  {author} {\bibinfo {author} {\bibfnamefont {A.~R.}\ \bibnamefont
  {Pries}}, \bibinfo {author} {\bibfnamefont {K.}~\bibnamefont {Ley}}, \bibinfo
  {author} {\bibfnamefont {M.}~\bibnamefont {Claassen}}, \ and\ \bibinfo
  {author} {\bibfnamefont {P.}~\bibnamefont {Gaehtgens}},\ }\bibfield  {title}
  {\enquote {\bibinfo {title} {Red cell distribution at microvascular
  bifurcations},}\ }\href@noop {} {\bibfield  {journal} {\bibinfo  {journal}
  {Microvasc Res}\ }\textbf {\bibinfo {volume} {38}},\ \bibinfo {pages}
  {81--101} (\bibinfo {year} {1989})}\BibitemShut {NoStop}%
\bibitem [{\citenamefont {Lahey}(1986)}]{lahey1986}%
  \BibitemOpen
  \bibfield  {author} {\bibinfo {author} {\bibfnamefont {R.~T.}\ \bibnamefont
  {Lahey}},\ }\bibfield  {title} {\enquote {\bibinfo {title} {Current
  understanding of phase separation mechanisms in branching conduits},}\
  }\href@noop {} {\bibfield  {journal} {\bibinfo  {journal} {Nucl. Eng.
  Design}\ }\textbf {\bibinfo {volume} {95}},\ \bibinfo {pages} {145--161}
  (\bibinfo {year} {1986})}\BibitemShut {NoStop}%
\bibitem [{\citenamefont {Azzopardi}(1993)}]{Azzopardi1993}%
  \BibitemOpen
  \bibfield  {author} {\bibinfo {author} {\bibfnamefont {B.~J.}\ \bibnamefont
  {Azzopardi}},\ }\bibfield  {title} {\enquote {\bibinfo {title} {T junctions
  as phase separators for gas liquid flows: possibilities and problems},}\
  }\href@noop {} {\bibfield  {journal} {\bibinfo  {journal} {Chem. Eng. Res.}\
  }\textbf {\bibinfo {volume} {71}},\ \bibinfo {pages} {273--281} (\bibinfo
  {year} {1993})}\BibitemShut {NoStop}%
\bibitem [{\citenamefont {Azzopardi}(1999)}]{Azzopardi1999}%
  \BibitemOpen
  \bibfield  {author} {\bibinfo {author} {\bibfnamefont {B.~J.}\ \bibnamefont
  {Azzopardi}},\ }\bibfield  {title} {\enquote {\bibinfo {title} {Phase
  separation at t-junctions},}\ }\href@noop {} {\bibfield  {journal} {\bibinfo
  {journal} {Multiphase Sci. Technol.}\ }\textbf {\bibinfo {volume} {11}},\
  \bibinfo {pages} {223--329} (\bibinfo {year} {1999})}\BibitemShut {NoStop}%
\bibitem [{\citenamefont {Azzopardi}\ and\ \citenamefont
  {Hervieu}(1994)}]{Azzopardi1994}%
  \BibitemOpen
  \bibfield  {author} {\bibinfo {author} {\bibfnamefont {B.~J.}\ \bibnamefont
  {Azzopardi}}\ and\ \bibinfo {author} {\bibfnamefont {E.}~\bibnamefont
  {Hervieu}},\ }\bibfield  {title} {\enquote {\bibinfo {title} {Phase
  separation at junctions},}\ }\href@noop {} {\bibfield  {journal} {\bibinfo
  {journal} {Multiphase Sci. Technol.}\ }\textbf {\bibinfo {volume} {8}},\
  \bibinfo {pages} {645--714} (\bibinfo {year} {1994})}\BibitemShut {NoStop}%
\bibitem [{\citenamefont {Shoham}\ \emph {et~al.}(1987)\citenamefont {Shoham},
  \citenamefont {Brill},\ and\ \citenamefont {Taitel}}]{shoham1987}%
  \BibitemOpen
  \bibfield  {author} {\bibinfo {author} {\bibfnamefont {O.}~\bibnamefont
  {Shoham}}, \bibinfo {author} {\bibfnamefont {J.~P.}\ \bibnamefont {Brill}}, \
  and\ \bibinfo {author} {\bibfnamefont {Y.}~\bibnamefont {Taitel}},\
  }\bibfield  {title} {\enquote {\bibinfo {title} {Two-phase flow splitting in
  a tee junction - experiment and modeling},}\ }\href@noop {} {\bibfield
  {journal} {\bibinfo  {journal} {Chem. Eng. Sci.}\ }\textbf {\bibinfo {volume}
  {42}},\ \bibinfo {pages} {2667--2676} (\bibinfo {year} {1987})}\BibitemShut
  {NoStop}%
\bibitem [{\citenamefont {Yang}\ \emph {et~al.}(2006)\citenamefont {Yang},
  \citenamefont {Azzopardi},\ and\ \citenamefont {Belghazi}}]{yang2006}%
  \BibitemOpen
  \bibfield  {author} {\bibinfo {author} {\bibfnamefont {L.}~\bibnamefont
  {Yang}}, \bibinfo {author} {\bibfnamefont {B.~J.}\ \bibnamefont {Azzopardi}},
  \ and\ \bibinfo {author} {\bibfnamefont {A.}~\bibnamefont {Belghazi}},\
  }\bibfield  {title} {\enquote {\bibinfo {title} {Phase separation of
  liquid-liquid two-phase flow at a t-junction},}\ }\href@noop {} {\bibfield
  {journal} {\bibinfo  {journal} {AICHE Journal}\ }\textbf {\bibinfo {volume}
  {52}},\ \bibinfo {pages} {141--149} (\bibinfo {year} {2006})}\BibitemShut
  {NoStop}%
\bibitem [{\citenamefont {Yang}\ and\ \citenamefont
  {Azzopardi}(2007)}]{yang2007}%
  \BibitemOpen
  \bibfield  {author} {\bibinfo {author} {\bibfnamefont {L.}~\bibnamefont
  {Yang}}\ and\ \bibinfo {author} {\bibfnamefont {B.~J.}\ \bibnamefont
  {Azzopardi}},\ }\bibfield  {title} {\enquote {\bibinfo {title} {Phase split
  of liquid-liquid two-phase flow at a horizontal t-junction},}\ }\href@noop {}
  {\bibfield  {journal} {\bibinfo  {journal} {Int. J. Multiphase Flow}\
  }\textbf {\bibinfo {volume} {33}},\ \bibinfo {pages} {207--216} (\bibinfo
  {year} {2007})}\BibitemShut {NoStop}%
\bibitem [{\citenamefont {Wang}\ \emph {et~al.}(2008)\citenamefont {Wang},
  \citenamefont {Wu}, \citenamefont {Zheng}, \citenamefont {Guo}, \citenamefont
  {Zhang},\ and\ \citenamefont {Tang}}]{wang2008}%
  \BibitemOpen
  \bibfield  {author} {\bibinfo {author} {\bibfnamefont {L.-Y.}\ \bibnamefont
  {Wang}}, \bibinfo {author} {\bibfnamefont {Y.-X.}\ \bibnamefont {Wu}},
  \bibinfo {author} {\bibfnamefont {Z.-C.}\ \bibnamefont {Zheng}}, \bibinfo
  {author} {\bibfnamefont {J.}~\bibnamefont {Guo}}, \bibinfo {author}
  {\bibfnamefont {J.}~\bibnamefont {Zhang}}, \ and\ \bibinfo {author}
  {\bibfnamefont {C.}~\bibnamefont {Tang}},\ }\bibfield  {title} {\enquote
  {\bibinfo {title} {Oil-water two-phase flow inside t-junction},}\ }\href@noop
  {} {\bibfield  {journal} {\bibinfo  {journal} {Journal of Hydrodynamics}\
  }\textbf {\bibinfo {volume} {20}},\ \bibinfo {pages} {147--153} (\bibinfo
  {year} {2008})}\BibitemShut {NoStop}%
\bibitem [{\citenamefont {Hongliang}\ \emph {et~al.}(2009)\citenamefont
  {Hongliang}, \citenamefont {Huanxin}, \citenamefont {Junlong}, \citenamefont
  {Hongge},\ and\ \citenamefont {Yunpeng}}]{hongliang2009}%
  \BibitemOpen
  \bibfield  {author} {\bibinfo {author} {\bibfnamefont {L.}~\bibnamefont
  {Hongliang}}, \bibinfo {author} {\bibfnamefont {C.}~\bibnamefont {Huanxin}},
  \bibinfo {author} {\bibfnamefont {X.}~\bibnamefont {Junlong}}, \bibinfo
  {author} {\bibfnamefont {T.}~\bibnamefont {Hongge}}, \ and\ \bibinfo {author}
  {\bibfnamefont {H.}~\bibnamefont {Yunpeng}},\ }\bibfield  {title} {\enquote
  {\bibinfo {title} {Refrigerant flow distributary disequilibrium caused by
  configuration of two phase fluid pipe network},}\ }\href@noop {} {\bibfield
  {journal} {\bibinfo  {journal} {Energy Conv. Manag.}\ }\textbf {\bibinfo
  {volume} {50}},\ \bibinfo {pages} {730--738} (\bibinfo {year}
  {2009})}\BibitemShut {NoStop}%
\bibitem [{\citenamefont {Minzer}\ \emph {et~al.}(2006)\citenamefont {Minzer},
  \citenamefont {Barnea},\ and\ \citenamefont {Taitel}}]{minzer2006}%
  \BibitemOpen
  \bibfield  {author} {\bibinfo {author} {\bibfnamefont {U.}~\bibnamefont
  {Minzer}}, \bibinfo {author} {\bibfnamefont {D.}~\bibnamefont {Barnea}}, \
  and\ \bibinfo {author} {\bibfnamefont {Y.}~\bibnamefont {Taitel}},\
  }\bibfield  {title} {\enquote {\bibinfo {title} {Flow rate distribution in
  evaporating parallel pipes - modeling and experiment},}\ }\href@noop {}
  {\bibfield  {journal} {\bibinfo  {journal} {Chem. Eng. Sci}\ }\textbf
  {\bibinfo {volume} {61}},\ \bibinfo {pages} {7249--7259} (\bibinfo {year}
  {2006})}\BibitemShut {NoStop}%
\bibitem [{\citenamefont {Lide}(2007)}]{RLide2007}%
  \BibitemOpen
  \bibfield  {author} {\bibinfo {author} {\bibfnamefont {D.~R.}\ \bibnamefont
  {Lide}},\ }\href@noop {} {\emph {\bibinfo {title} {CRC Handbook of Chemistry
  and Physics}}}\ (\bibinfo  {publisher} {CRC Press},\ \bibinfo {year}
  {2007})\BibitemShut {NoStop}%
\bibitem [{\citenamefont {Yih}(1967)}]{Yih1967}%
  \BibitemOpen
  \bibfield  {author} {\bibinfo {author} {\bibfnamefont {C.~S.}\ \bibnamefont
  {Yih}},\ }\bibfield  {title} {\enquote {\bibinfo {title} {Instability due to
  viscosity stratification},}\ }\href@noop {} {\bibfield  {journal} {\bibinfo
  {journal} {J. Fluid Mech.}\ }\textbf {\bibinfo {volume} {27}},\ \bibinfo
  {pages} {337--352} (\bibinfo {year} {1967})}\BibitemShut {NoStop}%
\bibitem [{\citenamefont {Talon}\ and\ \citenamefont
  {Meiburg}(2011)}]{Talon2011}%
  \BibitemOpen
  \bibfield  {author} {\bibinfo {author} {\bibfnamefont {L.}~\bibnamefont
  {Talon}}\ and\ \bibinfo {author} {\bibfnamefont {E.}~\bibnamefont
  {Meiburg}},\ }\bibfield  {title} {\enquote {\bibinfo {title} {Plane
  poiseuille flow of miscible layers with different viscosities: instabilities
  in the stokes flow regime},}\ }\href@noop {} {\bibfield  {journal} {\bibinfo
  {journal} {J. Fluid Mech.}\ }\textbf {\bibinfo {volume} {686}},\ \bibinfo
  {pages} {484--506} (\bibinfo {year} {2011})}\BibitemShut {NoStop}%
\bibitem [{\citenamefont {Gemmell}\ and\ \citenamefont
  {Epstein}(1962)}]{Gemmell:1962p5510}%
  \BibitemOpen
  \bibfield  {author} {\bibinfo {author} {\bibfnamefont {A.~R.}\ \bibnamefont
  {Gemmell}}\ and\ \bibinfo {author} {\bibfnamefont {N.}~\bibnamefont
  {Epstein}},\ }\bibfield  {title} {\enquote {\bibinfo {title} {Numerical
  analysis of stratified laminar flow of two immiscible newtonian liquids in a
  circular pipe},}\ }\href@noop {} {\bibfield  {journal} {\bibinfo  {journal}
  {The Canadian Journal of Chemical Engineering}\ }\textbf {\bibinfo {volume}
  {40}},\ \bibinfo {pages} {215--224} (\bibinfo {year} {1962})}\BibitemShut
  {NoStop}%
\end{thebibliography}%

\end{document}